\documentclass[reprint,prd,longbibliography,nofootinbib,amsmath,amssymb,aps]{revtex4-2}
\usepackage{graphicx}
\usepackage{dcolumn}
\usepackage{bm}
\usepackage{mathtools}
\usepackage{csquotes}
\usepackage{hyperref}
\usepackage{float}
\usepackage{bigints}
\usepackage{color}
\usepackage{xspace}
\usepackage{subcaption}
\usepackage{multirow}
\usepackage{booktabs}

\definecolor{blue}{rgb}{0.15, 0.38, 0.61}
\definecolor{red}{rgb}{0.71, 0.25, 0.15}
\definecolor{purple}{rgb}{0.51, 0.45, 0.65}
\definecolor{orange}{rgb}{0.81, 0.45, 0.35}

\begin{document}

\newcommand{\physrep}{{Phys. Rep.}}
\newcommand{\AAA}{\boldsymbol{A}}
\newcommand{\BB}{\boldsymbol{B}}
\newcommand{\JJ}{\boldsymbol{J}}
\newcommand{\EE}{\boldsymbol{E}}
\newcommand{\UU}{\boldsymbol{U}}
\newcommand{\kk}{\boldsymbol{k}}
\newcommand{\xx}{\boldsymbol{x}}
\newcommand{\rr}{\boldsymbol{r}}
\newcommand{\Bnabla}{\boldsymbol{\nabla}}
\newcommand{\ii}{\mathrm{i}}
\newcommand{\bra}[1]{\langle #1\rangle}
\newcommand{\bbra}[1]{\left\langle #1\right\rangle}
\newcommand{\eqss}[2]{(\ref{#1})--(\ref{#2})}
\newcommand{\EEq}[1]{Equation~(\ref{#1})}
\newcommand{\Eq}[1]{Eq.~(\ref{#1})}
\newcommand{\Eqs}[2]{Eqs.~(\ref{#1}) and~(\ref{#2})}
\newcommand{\Eqss}[2]{Eqs.~(\ref{#1})--(\ref{#2})}
\def\EM{E_{\rm M}}
\def\EA{E_{\rm A}}
\def\HM{H_{\rm M}}
\def\Lu{\mbox{\rm Lu}}
\def\MATINS{\texttt{MATINS}\xspace}
\def \clara#1{{\color{red}CD: #1}}

\title{Magnetar field dynamics driven by chiral anomalies without magnetic helicity}

\author{Clara Dehman$^{1}$}
\email{clara.dehman@ua.es}

\affiliation{$^{1}$ Departament de Física, Universitat d'Alacant, 03690 Alicante, Spain}

\newcommand{\jcap}{{JCAP}}
\newcommand{\aap}{{A\&A}}
\newcommand{\mnras}{{MNRAS}}
\newcommand{\apjl}{{ApJL,}}
\newcommand{\apjs}{{ApJS,}}
\newcommand{\ssr}{Space Sci. Rev.,}

\begin{abstract}
The chiral magnetic effect (CME), arising from the chiral anomaly and enabling a mutual conversion between magnetic topology and fermionic chirality, is a key mechanism in magnetar field evolution. Previous work \citep{DehmanPons2025} demonstrated that the CME can efficiently generate dipolar fields ($B_{\rm dip} \gtrsim 10^{14}\,\mathrm{G}$), consistent with magnetar timing measurements, provided that the initial magnetic field carries net helicity. However, whether neutron stars are born with magnetic helicity remains uncertain. In this work, we investigate the CME across a range of initial helicity configurations, including non-helical initial conditions. We find that the CME efficiently generates magnetar-strength dipoles on timescales of decades, independently of the initial helicity content. The chiral instability is driven by localized helical structures that induce a residual chiral asymmetry and is primarily governed by the maximum chiral chemical potential, requiring $\mu_5^{\rm max} \gtrsim \mathrm{few}\times10^{-11}\,\mathrm{MeV}$ for onset in the magnetar regime. Our results further show that these dipoles may either remain stable and subsequently evolve through standard Ohmic decay, or become unstable if they acquire sufficient helicity, in which case they decay through the chiral anomaly, transferring energy to less helical modes. This outcome depends sensitively on the initial helicity distribution. These findings extend the applicability of the CME to more realistic magnetic-field configurations and underscore the importance of the helicity distribution at birth---a quantity that remains poorly constrained in newborn neutron stars, yet is crucial for determining their magnetic evolution and the emergence of magnetars.
\end{abstract}

\maketitle

\section{Introduction}

The transport of electric charge driven by quantum anomalies in chiral fermion systems has recently become a focal point of interest, particularly in the study of neutron stars and magnetars \citep{ohnishi2014,DehmanPons2025}. This surge of attention reflects the possibility of accessing a distinct class of macroscopic quantum behavior that can directly influence the evolution of astrophysical magnetic fields. Conventional macroscopic quantum phenomena—such as superfluidity and superconductivity—emerge through symmetry breaking and are described by a local order parameter, for instance the Cooper-pair density \citep{Poniatowski2019}. By contrast, anomaly-induced effects in chiral systems originate from global topological properties and do not rely on symmetry breaking.

One of the most prominent manifestations of quantum anomalies is the chiral magnetic effect (CME), whereby a chirality imbalance among charged fermions induces an electric current parallel to an external magnetic field $\mathbf{B}$ via the Adler–Bell–Jackiw anomaly \citep{adler1969,bell1969}. The required chiral imbalance is tied, through the Atiyah–Singer index theorem \citep{Freed2021}, to the global topological structure of the gauge field. The CME provides a direct coupling between microscopic quantum anomalies and macroscopic magnetic-field evolution. In this framework, the induced current enters Maxwell’s equations as a source term, so that a current aligned with the magnetic field twists magnetic flux and generates non-zero magnetic helicity. This coupling is reciprocal: magnetic helicity can in turn source chiral asymmetry as the field relaxes and untwists \citep{boyarsky2012,DehmanPons2025}, thereby establishing a feedback loop between fermionic chirality and magnetic topology.

The significance of this mechanism has long been questioned due to efficient, temperature-dependent chirality-flipping processes induced by finite fermion masses, in particular the electron mass in astrophysical plasmas, which tend to damp the chiral imbalance \citep{grabowska2015,sigl2016,kaplan2017} and thereby limit its impact on magnetic-field evolution. While this suppression is effective in short-lived environments such as proto-neutron stars and the early Universe \citep{sigl2016,skoutnev2026}, recent three-dimensional magneto-thermal simulations \citep{DehmanPons2025} show that, in long-lived neutron stars, the CME can still play a significant role even when such processes are fully taken into account. In particular, Ref.~\cite{DehmanPons2025} showed that an initially small-scale helical magnetic field\footnote{A magnetic field configuration with non-zero magnetic helicity, reflecting a net global twist, writhe, or linkage of its field lines.} can, through the chiral anomaly, generate a chiral asymmetry between left- and right-handed electrons that is sustained for hundreds of years. This asymmetry drives an inverse transfer of magnetic energy toward large scales, leading to the emergence of magnetar-strength dipolar components on a comparable (century-long) timescale — without requiring a pre-existing chiral imbalance.

This result is particularly significant for neutron stars, whose magnetic-field configuration at birth is expected to be dominated by small-scale structures rather than a large-scale dipole. Magnetohydrodynamic (MHD) simulations indicate that the magnetic field is amplified during the proto-neutron star phase by a turbulent dynamo that can generate magnetar-level magnetic energies; however, most of this energy is stored in small-scale, non-axisymmetric, and predominantly toroidal components, leaving only a weak large-scale dipole \citep{balbus1991, obergaulinger2014, aloy2021, reboul2021, masada2022, matsumoto2022, barrere2025}. Timing observations show that magnetars are slow rotators, implying the presence of strong large-scale surface dipolar fields of order $10^{14}$~G. Three-dimensional long-term magneto-thermal evolution studies show that such initial magnetic field configurations cannot reproduce all magnetars' properties when only Ohmic dissipation and Hall drift are considered \citep{dehman2023b, igoshev2025, dehmanbrandenburg2025}, although such configurations may still be relevant for other neutron star classes, such as Central Compact Objects or low-field magnetars. This discrepancy suggests that an additional mechanism capable of transferring magnetic energy from small to large scales is required, making the CME a natural candidate.

The efficiency of the CME as a driver of inverse magnetic-energy transfer relies on the key assumption that the neutron star is born with an initially helical magnetic field \citep{DehmanPons2025}. However, the magnetic helicity content at birth remains largely unconstrained, and whether newborn neutron stars carry a net helicity remains unclear, with theoretical arguments supporting both possibilities. Standard MHD simulations preserve reflection symmetry, so that an initially vanishing helicity leads to the development of mirror-symmetric structures with opposite handedness \citep{Woltjer1958, taylor1974, bodo2017}, allowing magnetic energy to grow while the net helicity remains negligible \citep{brandenburg2005b}. By contrast, net helicity may be generated if the CME operates during core-collapse supernovae: as the initial chiral asymmetry is gradually erased while the system approaches chemical equilibrium, the anomaly converts it into magnetic helicity \citep{matsumoto2022}. This process, however, is strongly suppressed by spin-flip reactions, which rapidly damp the chiral imbalance through a channel unrelated to the anomaly, thereby limiting the amount of net helicity that can be generated.

Motivated by these uncertainties, we explore a range of magnetic field configurations representative of newborn neutron stars, systematically varying both the initial helicity content and its spatial distribution. Our goal is to assess whether the CME remains active when the initial magnetic field is non-helical, as may be expected at neutron star birth, and to identify which configurations can evolve into stable and strong dipolar fields characteristic of magnetars.

The structure of the paper is as follows. Section \ref{sec: formalism} introduces the theoretical framework employed in this work. Section \ref{sec: numerical setup} details the numerical setup and initial conditions. The results are presented in Section \ref{sec: results} and discussed in Section \ref{sec: discussion}.

\section{Magnetic field formalism}
\label{sec: formalism}

Magnetic helicity is a fundamental topological invariant, expressible in terms of the Chern–Simons three-form, and provides a quantitative measure of the twist, writhe, and linkage of magnetic-field lines \citep{moffatt1969, berger1984, arnold1992, Pevtsov2014}. It remains nearly conserved even in non-ideal, high–magnetic-Reynolds-number regimes, as demonstrated in studies of magnetic reconnection in the solar atmosphere \citep{Pariat2015}, and is increasingly recognized as a key factor in neutron-star magnetic field evolution \citep{brandenburg2020, dehmanbrandenburg2025, DehmanPons2025}. It is defined as
\begin{equation}
    \chi_M = \int_V \AAA \cdot \BB \, dV , 
    \label{eq: A.B}
\end{equation}
where $\AAA$ is the magnetic vector potential, $\BB = \Bnabla \times \AAA$ is the magnetic field, and $V$ denotes the integration volume.

The local time evolution of the magnetic helicity density, $\AAA \cdot \BB$, is given by \citep{biskamp1997}
\begin{eqnarray}
\frac{\partial (\AAA \cdot \BB)}{\partial t} 
&=& - 2 c  \EE \cdot \BB -  c \Bnabla \cdot \left( \EE \times \AAA\right) ,
 \label{eq: magnetic helicity evolution}
\end{eqnarray}
where $c$ is the speed of light and $\EE$ is the electric field.
The second term, $-c\,\Bnabla\cdot(\EE\times\AAA)$, represents the helicity flux --- it describes how helicity is shuffled between different regions of the star, even in the absence of any net transport across the stellar boundary. Only upon volume integration over the full stellar interior does this term reduce to a boundary contribution, which vanishes for an isolated neutron star.
The first term, $-2c \, \EE \cdot \BB$, governs the evolution of magnetic helicity density and simultaneously drives the generation of chiral asymmetry through the chiral anomaly \citep{treiman1986}. This links electromagnetic fields to fermionic chirality and gives rise to the evolution equation for the chiral number density, $n_5 \equiv n_R - n_L$, \citep{adler1969, kamada2023, DehmanPons2025}:
\begin{equation}
    \frac{\partial n_5}{\partial t} =  \frac{2 \alpha }{\pi \hslash} \EE \cdot \BB - n_5 \Gamma_f.
    \label{eq: n5}
\end{equation}
Equation~\ref{eq: n5} contains both a source and a sink term. The $\EE \cdot \BB$ contribution can act as either, depending on its sign and thus on whether magnetic field lines are effectively twisted or untwisted. In contrast, the reaction rate $\Gamma_f$ accounts for spin-flip interactions arising from electromagnetic effects and the finite electron mass; it always acts as a sink, reducing the efficiency of the CME and suppressing the chiral magnetic instability (CMI) \citep{grabowska2015, sigl2016, kaplan2017, kamada2023}. The magnitude of the spin-flip rate casts significant doubt on the relevance of the CME in short-lived environments such as proto-neutron stars and the early Universe \citep{skoutnev2026}. 
Nonetheless, chiral asymmetries have been shown to persist for centuries in long-lived neutron stars despite this suppression \citep{DehmanPons2025}, enabling a sustained transfer of magnetic energy from small to large scales. For a simplified description of how the chiral anomaly operates, see Section~\ref{sec: chiral anomaly}.

\Eq{eq: magnetic helicity evolution} and \Eq{eq: n5}, when combined and integrated over the volume, yield a generalized chiral–magnetic helicity balance law \cite{treiman1986,boyarsky2012,rogachevskii2017,DehmanPons2025}:
 \begin{equation}
    \frac{d}{dt} \left( Q_5 + \frac{\alpha }{\pi \hslash c} \chi_m  \right) + {\Gamma_5} = 0,
    \label{eq: modified helicity conservation}
\end{equation}
where $Q_5 = \int n_5 \, dV$ is the total axial charge, quantifying the global imbalance between left- and right-handed fermions, and ${\Gamma_5} = \int n_5 \, \Gamma_f \, dV$ is the total spin-flip dissipation rate. The appearance of the sink term $\Gamma_5$ implies that this quantity is not strictly conserved.

The resulting local imbalance between left- and right-handed electrons induces an electric current parallel to the magnetic field through the Adler–Bell–Jackiw anomaly \citep{adler1969, bell1969}, giving rise to an additional term in Maxwell’s equations \citep{vilenkin1980,ohnishi2014}:
\begin{equation}
    \JJ_5 = \frac{\alpha \mu_5}{\pi \hslash } \BB, 
    \label{eq: J5}
\end{equation}
where $\mu_5 \equiv \mu_R - \mu_L$ is the chiral chemical potential, and $\alpha = e^2 / (\hslash c)$ is the fine-structure constant, with $e$ the elementary charge and $\hslash$ the reduced Planck constant. Gaussian units are used throughout this section.

In this work, we focus on the CME in the neutron star crust, where the spin-flip rate is given by $\Gamma_f = 4\alpha / (3\pi \sigma_e)$, with $\sigma_e$ denoting the electrical conductivity. This choice is motivated by our previous results \citep{DehmanPons2025}, which identified the crust as the region where the CME remains efficient despite spin-flip suppression. In contrast, the neutron star core---which contains most of the stellar volume---is characterized by much stronger spin-flip damping, primarily due to electron scattering off magnetized neutron vortices in the superfluid \citep{Feibelman1971, Alpar1984}. These processes occur on significantly shorter timescales than in the crust and are therefore expected to strongly suppress any chiral asymmetry. As a result, it remains unclear whether a residual chiral asymmetry could survive long enough to influence magnetic field evolution in the core, a question that warrants further investigation in future work.

Under these conditions, the magnetic field evolution in the crust is governed by the modified induction equation, which incorporates Ohmic dissipation, Hall drift, and the chiral magnetic contribution \citep{DehmanPons2025}:
\begin{eqnarray}
\frac{\partial \BB}{\partial t} =  - \Bnabla \times \left[  \eta \left( \Bnabla \times \BB  - k_5  \BB \right)+ f_h \left(\Bnabla \times \BB \right)\times \BB
     \right]. ~
    \label{Eq:CME}
\end{eqnarray}
The first term on the right-hand side represents Ohmic dissipation, where the magnetic diffusivity is defined as $\eta \equiv c^{2}/(4\pi\sigma_e)$. It leads to the decay of the magnetic field and is most effective at small spatial scales.
The second term corresponds to the CME, with the chiral wavenumber defined as $k_5 = 4\alpha \mu_5/(\hbar c)$. This term redistributes magnetic energy across spatial scales, primarily transferring it from small-scale to large-scale magnetic structures.
The last term represents the Hall drift, where $f_h = c/(4\pi e n_e)$ is the Hall prefactor, with $n_e$ the electron number density. The Hall drift is known to induce a direct cascade, transferring magnetic energy from large-scale structures to smaller scales, where it is more efficiently dissipated by Ohmic diffusion \citep{pons2007, dehman2023b}. In addition, in the presence of helical (or partially helical) magnetic fields, it can also drive an inverse cascade \citep{brandenburg2020, dehmanbrandenburg2025}. In this regime, the approximate conservation of magnetic helicity implies that, as magnetic energy is dissipated, the characteristic wavenumber shifts toward smaller values, corresponding to larger spatial scales. Consequently, the inverse cascade proceeds on the magnetic dissipation timescale.
For a comprehensive overview of magneto-thermal evolution in neutron stars, we refer the reader to Ref.~\citep{pdv2025}.

\begin{figure*}[htbp]
\centering
\begin{subfigure}{0.32\textwidth}
    \centering
       \caption{$B \gtrsim B_{\rm QED}$} \includegraphics[width=\linewidth]{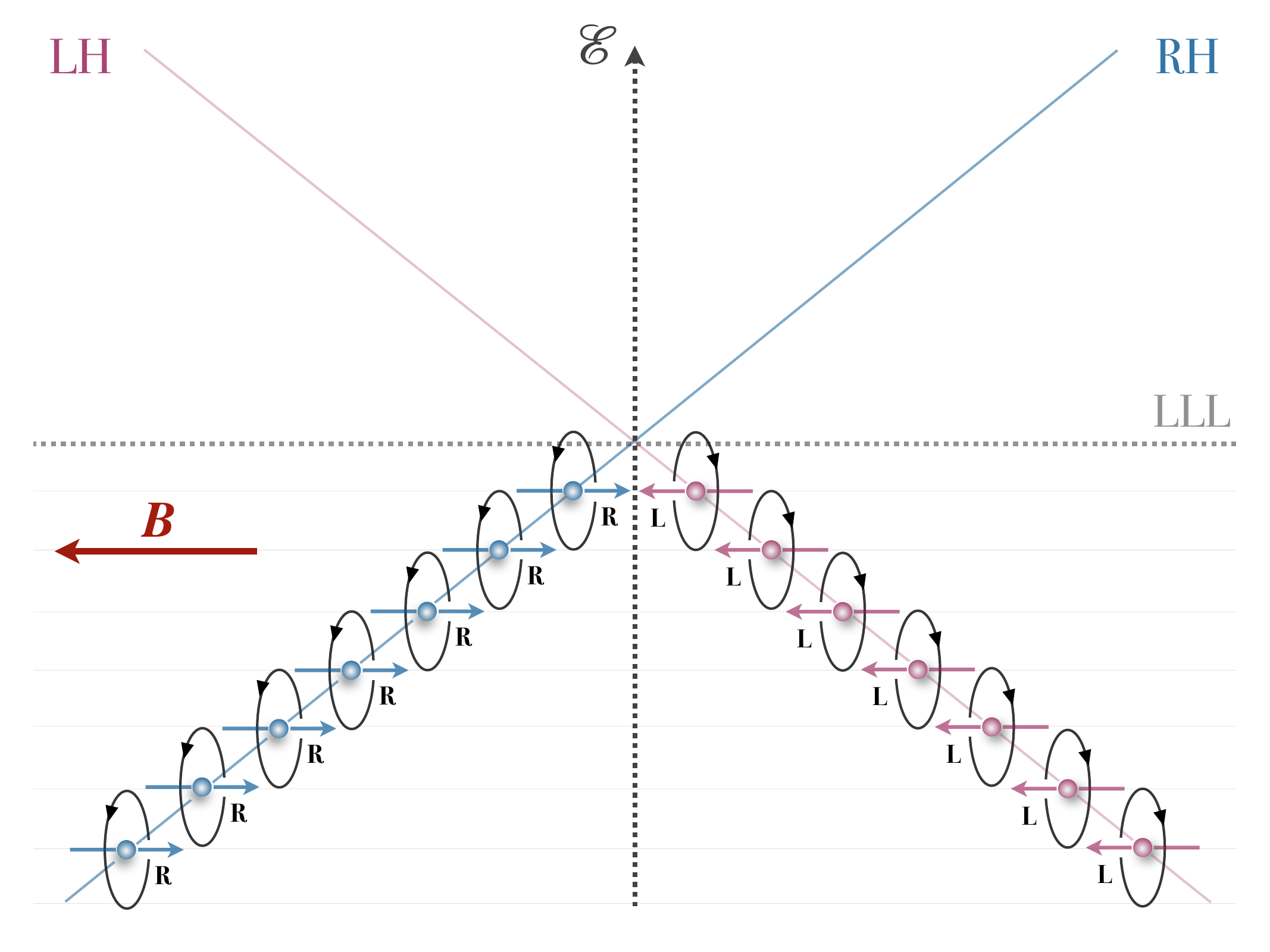}
    \label{fig:a}
\end{subfigure}
\hfill
\begin{subfigure}{0.32\textwidth}
    \centering
        \caption{$\EE \cdot \BB > 0$}
\includegraphics[width=\linewidth]{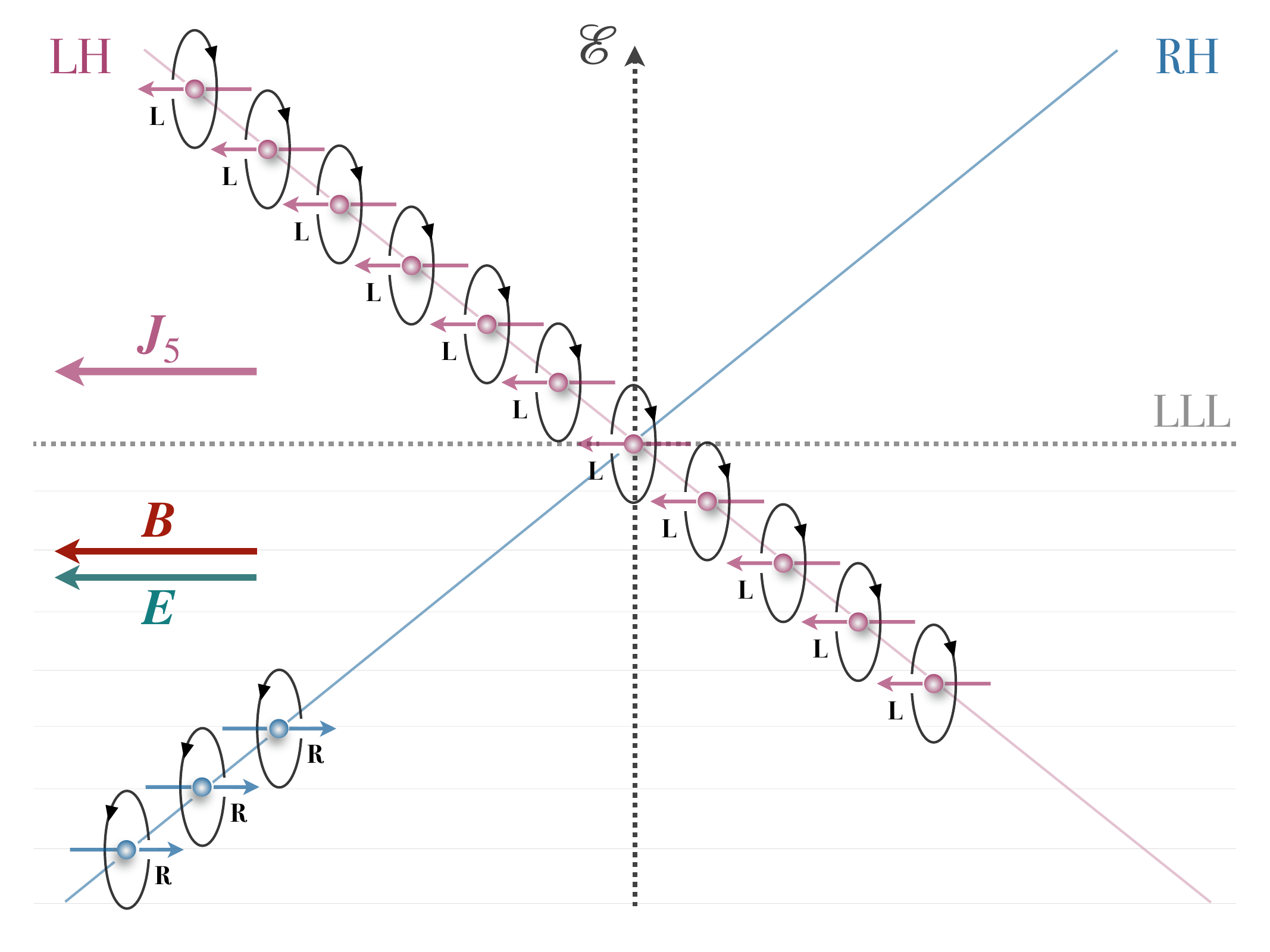}
    \label{fig:b}
\end{subfigure}
\hfill
\begin{subfigure}{0.32\textwidth}
    \centering
        \caption{$\Gamma_f \neq 0$}
        \includegraphics[width=\linewidth]{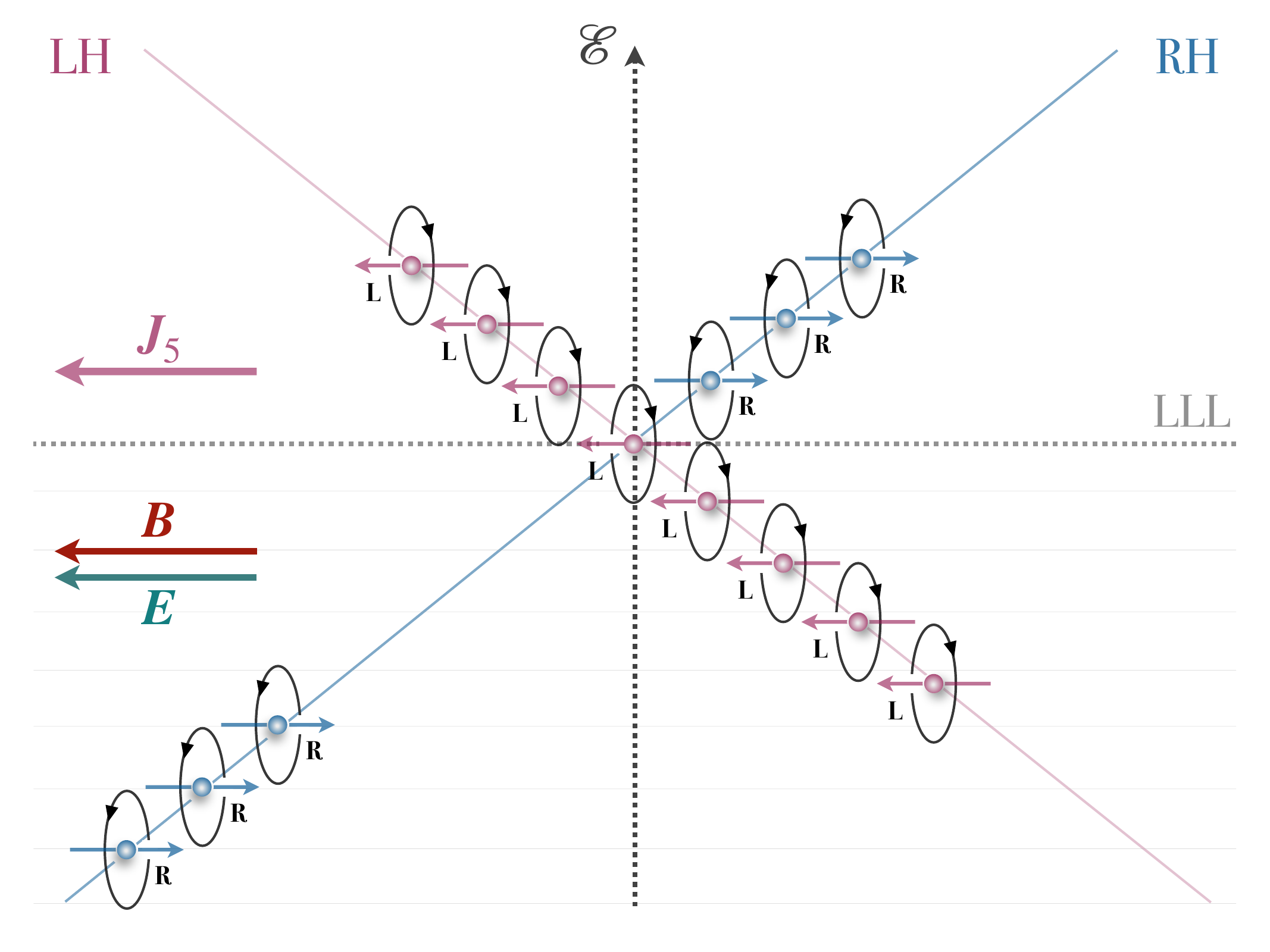}
    \label{fig:c}
\end{subfigure}
\caption{Illustration of the chiral anomaly in a strong magnetic field regime, $B \gtrsim B_{\rm QED}$, oriented along the $x$-direction. Right-handed and left-handed electron momenta are shown in blue and pink, respectively. 
\textit{Panel (a):} Electrons are constrained to propagate along the magnetic field. Right-handed electrons move opposite to $\BB$, while left-handed electrons move along $\BB$. The system initially contains equal populations of both chiralities. 
\textit{Panel (b):} As magnetic helicity relaxes, a positive $\EE\cdot \BB$ is induced. Right-handed electrons are converted into left-handed ones via momentum reversal, and the resulting left-handed electrons populate higher Landau levels. Consequently, a net chiral asymmetry is generated.
\textit{Panel (c):} Chirality-flipping processes acting on electrons in higher Landau levels relax the chiral imbalance on short timescales.
}
\label{fig:chiral anomaly}
\end{figure*}

\subsection{Quasi-equilibrium and astrophysical timescales}
\label{sec: quasi-equilibrium}
When spin-flip reactions are included, their rates are extremely large—of order $10^{15}-10^{17}$ s$^{-1}$ under typical neutron-star conditions during the first few centuries. Because these rates exceed macroscopic magnetic-evolution timescales by many orders of magnitude, the system remains in a quasi-equilibrium state. In this regime—effectively always realized in astrophysical contexts—one can derive an explicit expression for $k_5$ as a function of the magnetic field and its helicity density \citep[see Section~II of][]{DehmanPons2025}:
\begin{equation}
    k_5 =  \frac{ k_B } {1 + \left(\dfrac{2 \mu_e^2 }{ m_e^2 c^4}\right) \dfrac{B^2_\mathrm{QED}}{ 3 \pi B^2} },
    \label{eq: k5}
\end{equation}
where 
\begin{equation}
    k_{B} \equiv \frac{ \left(\Bnabla\times \BB \right) \cdot \BB  }{ B^2} ,
    \label{eq: kb}
\end{equation}
and $B_\mathrm{QED} \equiv m_e^2 c^3 / (e \, \hslash) = 4.41 \times 10^{13}$\,G is the Schwinger critical field.
Equation~\eqref{eq: k5} shows that the chiral wavenumber $k_5$ is locally constrained by the characteristic wavenumber associated with currents aligned with the magnetic field, namely $k_{B}$. It also demonstrates that the CME operates efficiently only in the strong-field regime ($B \gtrsim B_\mathrm{QED}$), as realized in magnetars. This conclusion emerges naturally once spin-flip processes, often neglected in CME studies, are properly taken into account. 
In the quasi-equilibrium limit, \Eq{eq: magnetic helicity evolution} can be written as
\begin{equation}
    \frac{\partial (\AAA \cdot \BB)}{\partial t} 
=   \frac{- 2 \, \eta \,  \left( \Bnabla \times \BB \right)\cdot \BB}{1+ \dfrac{B^2}{B_{\rm sat}^2}}  -  c \Bnabla \cdot \left( \EE \times \AAA\right) , 
\label{eq: magnetic helicity evolution quasi-equilibrium}
\end{equation}
where $B_\mathrm{sat}$ is the characteristic magnetic field strength at which saturation sets in,
\begin{equation}
B_\mathrm{sat} \approx \sqrt{\frac{2}{3 \pi}} \frac{\mu_e }{m_e c^2} B_\mathrm{QED}.
    \label{eq: Bmax}
\end{equation}
and scales linearly with $\mu_e/(m_e c^2)$, which increases with density. This yields $B_\mathrm{sat} \sim 10^{14}\,\mathrm{G}$ near the surface and up to $\sim 5 \times 10^{15}\,\mathrm{G}$ in deeper layers, consistent with inferred magnetar field strengths.

The first term in \Eq{eq: magnetic helicity evolution quasi-equilibrium} directly couples the evolution of magnetic helicity density to the current helicity density through the magnetic diffusivity. The sign and magnitude of $(\Bnabla \times \BB)\cdot \BB$ determine whether helicity is generated or dissipated. The efficiency of this process, and thus of the CME (see \Eq{eq: modified helicity conservation}), varies strongly with radius due to the steep diffusivity gradient. In regions of higher diffusivity, such as the outer crust and the nuclear pasta layer, magnetic field structures relax more rapidly, facilitating a more efficient conversion between magnetic helicity and chiral asymmetry. Once $B$ approaches $B_{\rm sat}$, the denominator in \Eq{eq: magnetic helicity evolution quasi-equilibrium} increasingly suppresses this coupling, capping the efficiency of the conversion and thereby saturating the CME.

\subsection{A simplified framework for the chiral anomaly}
\label{sec: chiral anomaly}

In ultra-strong magnetic fields, well above the Schwinger QED critical value $B_\mathrm{QED} \equiv m_e^2 c^3 / (e\,\hbar) = 4.41 \times 10^{13}$~G, as expected in magnetar interiors, electron motion perpendicular to the magnetic field is quantized into discrete Landau levels \citep{Kostenko2018,Thompson2020}. Because of the high electron densities characteristic of neutron star interiors, many Landau levels remain populated even when $B \gg B_\mathrm{QED}$; occupation of only the lowest Landau level (LLL) is a good approximation in the comparatively low-density crust ($\rho \lesssim 10^9$~g~cm$^{-3}$), where the electron Fermi energy falls below the energy of the first excited Landau level \citep{Sharma2011}. Below we describe the chiral anomaly mechanism in this simplified LLL limit, since it captures the essential physics of the process. In the LLL, the transverse degrees of freedom carry zero energy, so that electrons can only propagate freely along the field direction, and only the electron spin state whose magnetic moment is aligned with $\BB$ is kinematically allowed \citep{Landau1991,Miransky2015}. Because every electron is forced into this single spin state, its helicity is fixed entirely by its direction of propagation along $\BB$: spin and momentum become locked to a single chirality. If $\BB$ points along the $x$-direction, left-handed electrons propagate in the positive $x$-direction, while right-handed electrons propagate in the negative $x$-direction \citep{Kharzeev2008}.

This effect is illustrated in panel (a) of Figure~\ref{fig:chiral anomaly}, where the magnetic field is assumed to be uniform and oriented along the $x$-direction, i.e., $\BB = B_x \,\hat{\boldsymbol{x}}$. Initially, equal numbers of left-handed and right-handed electrons are present, corresponding to neutron-star matter in chemical equilibrium, with no net chiral asymmetry.

\begin{figure}
\centering
\includegraphics[width=\linewidth]{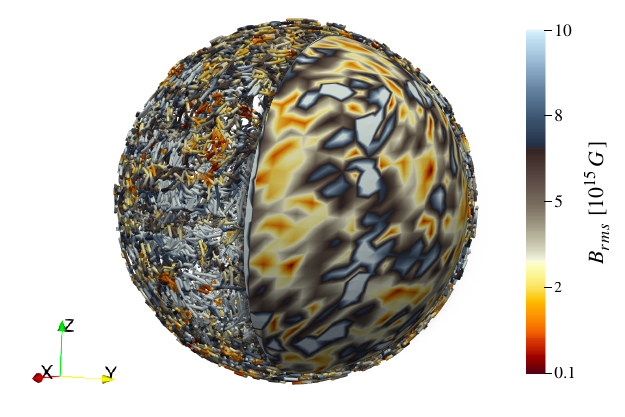} 
\caption{Schematic representation of the initial small-scale tangled magnetic field configuration.}
\label{fig: bfield}
\end{figure}

\begin{figure*}
\centering
\includegraphics[width=0.95\linewidth]{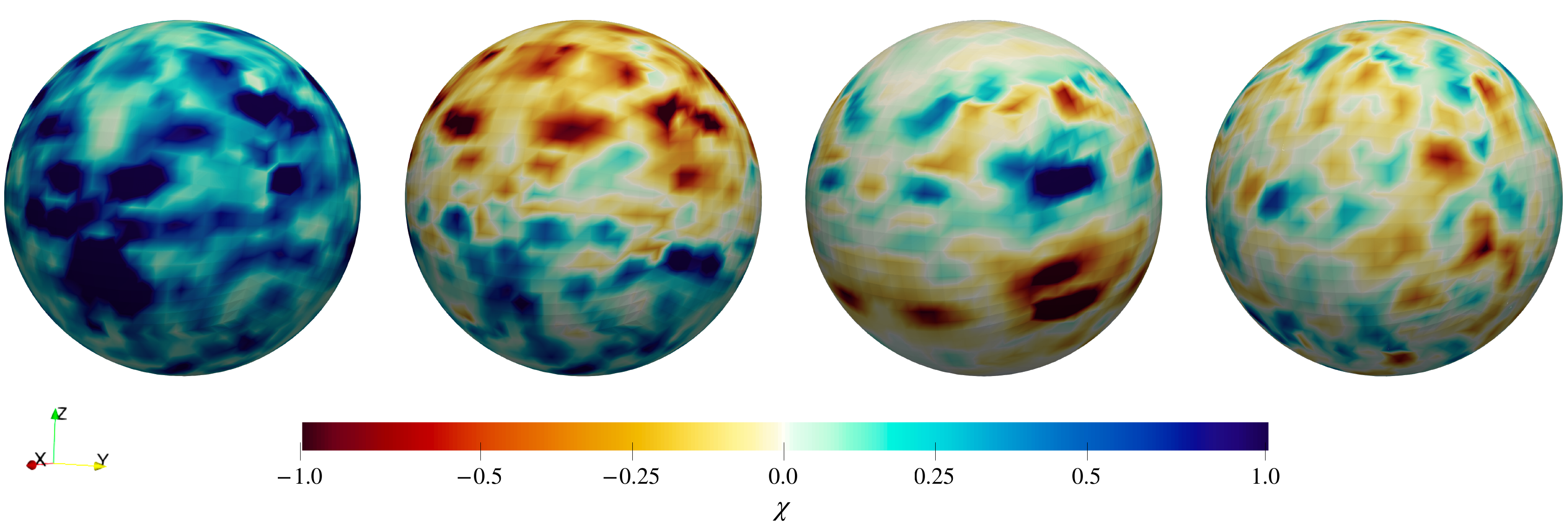} 
\caption{Schematic representation of $\chi$, the initial local magnetic helicity density, normalized by its maximum absolute value, for the \texttt{Monohel}, \texttt{Bihel}, \texttt{Mixhel}, and \texttt{Angfluc} setups, from left to right.}
\label{fig: initial helicity}
\end{figure*}

As magnetic field lines locally untwist, magnetic helicity decreases slightly owing to finite magnetic diffusivity, which induces a locally positive $\EE \cdot \BB$ term. As a result, some electrons undergo a chirality flip. Since true spin flips are energetically suppressed in a strong magnetic field, chirality can only change via a reversal of the electron momentum. For $\EE \cdot \BB > 0$, the induced electric field is locally aligned with the magnetic field, converting right-handed electrons into left-handed electrons via momentum reversal \citep{Kharzeev2008}; the newly created left-handed electrons that cannot be accommodated in the (fully occupied) LLL instead populate higher Landau levels. Although the total electron number, $N_e = N_R + N_L$, is conserved, the axial charge $N_5 = N_R - N_L$ is not: this generates an axial current $\JJ_5$ along the direction of $\BB$, reflecting the presence of the chiral (axial) anomaly. This mechanism is illustrated in panel (b) of Figure~\ref{fig:chiral anomaly}.

In addition to this anomalous generation of axial charge by a non-vanishing $\EE \cdot \BB$, chirality-flipping processes act to relax the resulting imbalance \citep{grabowska2015}. In the ultra-strong-field regime where electrons predominantly occupy the LLL, true spin-flip transitions cannot occur, since the LLL admits only a single spin polarization state \citep{Miransky2015}. Such transitions become possible only for electrons in higher Landau levels, where spin and momentum are no longer rigidly locked and chirality is not conserved \citep{Armitage2018}. Finite electron-mass effects and electromagnetic scattering processes then convert right- and left-handed electrons into one another, relaxing the axial charge $N_5$. These chirality-flipping processes occur on very short timescales and can therefore be treated as effectively instantaneous and local within the neutron star interior. Their role is illustrated in panel (c) of Figure~\ref{fig:chiral anomaly}.

It is important to note that an electric field anti-aligned with the magnetic field ($\EE \cdot \BB < 0$) corresponds to a system with a pre-existing chiral asymmetry, which is subsequently converted into magnetic helicity via the chiral anomaly relation (see Eq.~\ref{eq: magnetic helicity evolution}). Such a scenario is expected, for instance, in proto-neutron stars, where weak interactions preferentially produce left-handed electrons, leading to a chiral imbalance. As the neutron star forms and evolves toward chemical equilibrium, the populations of left- and right-handed electrons become comparable. During this evolution the system exhibits $\EE \cdot \BB < 0$, converting left-handed electrons into right-handed ones — corresponding to a transition from panel (b) to panel (a) in Figure~\ref{fig:chiral anomaly}.

\section{Numerical setup}
\label{sec: numerical setup}

To study the magnetic field evolution under the CME, we use the extended \MATINS code introduced in Ref.~\citep{DehmanPons2025}, which, in addition to solving the coupled induction and heat-diffusion equations, self-consistently evolves the chiral number density and incorporates spin-flip processes arising from the chiral anomaly. This implementation builds on the earlier version of \MATINS\footnote{\url{https://github.com/ice-csic-astroexotic/MATINS}} \citep{dehman2023,dehman2023b,ascenzi2024}, which did not account for the CME. Specifically, we solve the induction equation in Eq.~\ref{Eq:CME} while neglecting the Hall drift term, in order to isolate CME effects from Hall-driven evolution, since the Hall term also redistributes magnetic energy across spatial scales \citep{pons2007,dehman2023b,brandenburg2020,dehmanbrandenburg2025}; a discussion of its role is deferred to the end of Section~\ref{sec: results}. Since spin-flip reactions occur on timescales much shorter than those governing macroscopic magnetic evolution, the system is treated in quasi-equilibrium: the chiral wavenumber $k_5$ is obtained by solving the evolution of the chiral number density (Eq.~\ref{eq: n5}) in this limit, yielding Eqs.~\ref{eq: k5} and \ref{eq: kb}. Thermal evolution is computed by solving the heat-diffusion equation (as in Ref.~\cite{ascenzi2024}), with Joule heating modified to account for the CME contribution through the total dissipation $Q_{\mathrm{tot}} = \int \sigma_e \mathbf{E}^2 \, dV$, where, in the absence of the Hall term, the electric field satisfies $c \, \mathbf{E} = \eta (\nabla \times \mathbf{B} - k_5 \mathbf{B})$.

We model the entire neutron star volume but restrict magnetic field evolution to the crust. We impose potential-field (current-free) boundary conditions at a mass density of $\rho = 10^{10}\,\mathrm{g\,cm^{-3}}$, which defines the numerical surface of the star, and perfect-conductor conditions at the crust–core interface, thereby confining the magnetic field to the crust (see Ref.~\cite{dehman2023} for details on the magnetic boundary conditions). The temperature-dependent electrical conductivity is computed using the IOFFE codes\footnote{\href{http://www.ioffe.ru/astro/conduct/}{http://www.ioffe.ru/astro/conduct/}} \citep{potekhin2015}. The stellar background in \MATINS is constructed using zero-temperature equations of state (EOSs) from the CompOSE\footnote{\href{https://compose.obspm.fr/}{https://compose.obspm.fr/}} database. In this work, we adopt the BSk24 EOS \citep{pearson2018}, which has been shown to successfully explain the luminosities of thermally emitting neutron stars \citep{marino24}. We consider a canonical $1.4\,M_\odot$ neutron star, yielding $R = 12.4$\,km and a crust thickness of $0.86$\,km; \MATINS nonetheless allows exploration of alternative EOSs and stellar masses. For the blanketing envelope, we adopt the model of Ref.~\citep{potekhin2015}, which assumes a heavy-element composition (e.g., iron) and accounts for magnetic-field effects. For a discussion of the impact of different envelope compositions and envelope magnetization on the magneto-thermal evolution, we refer the reader to Ref.~\cite{dehman2023c}.

\begin{table*}[!ht]
\caption{Simulation data at different representative times, showing the magnetic field strength (mean and dipolar components), the chiral chemical potential (mean and maximum values), and the dimensionless fractional helicity, $\tilde{\chi}$.}
\centering
\setlength{\tabcolsep}{3pt} 
\begin{tabular}{lccccccccccc}
\hline
\noalign{\smallskip}
\multirow{2}{*}{Run} 
& \multicolumn{2}{c}{$B_{\rm rms}$ [G]} 
& \multicolumn{2}{c}{$B_{\rm dip}$ [G]} 
& \multicolumn{2}{c}{$\mu_5^{\rm avg}$ [MeV]} 
& \multicolumn{2}{c}{$\mu_5^{\rm max}$ [MeV]} 
& \multicolumn{2}{c}{$\tilde{\chi}$} \\
  & $t=0$ & $100$\,yr & $t=0$ & $100$\,yr & $t=0$ & $100$\,yr & $t=0$ & $100$\,yr &   $t=0$ & $1000$\,yr\\
\hline
\noalign{\smallskip}
\texttt{Monohel}  & $2.9\times10^{16}$ & $2.3\times10^{16}$ & $4.3\times10^{12}$ & $1.0\times10^{14}$ & $~~2\times10^{-12}$ & $~~1\times10^{-12}$ & $2\times10^{-11}$ & $1\times10^{-11}$ & $4 \times 10^{-6}$ &  $ 0.7 \,  \tilde{\chi}_{\rm Mo}^{t0}$\\ 
\noalign{\smallskip}
\texttt{Bihel}  & $3.0\times10^{16}$ & $2.0\times10^{16}$ & $4.1\times10^{12}$ & $1.6\times10^{14}$ & $-5\times10^{-14}$ & $-5\times10^{-14}$ & $8\times10^{-11}$ & $2\times10^{-11}$ & $3.5 \% \,  \tilde{\chi}_{\rm Mo}^{t0}$ & $ 1.7 \, \tilde{\chi}_{\rm Bi}^{t0}$\\ 
\noalign{\smallskip}
\texttt{Mixhel}  &  $3.1 \times 10^{16}$ & $1.8 \times 10^{16}$ & $4.0 \times 10^{12}$ & $1.4 \times 10^{14}$  & $ -2 \times 10^{-13}$ & $-1 \times 10^{-13} $ & $9 \times 10^{-11}$ & $2 \times 10^{-11}$ & $10\% \,  \tilde{\chi}_{\rm Mo}^{t0}$ &$ 1.6  \, \tilde{\chi}_{\rm Mi}^{t0}$\\
\noalign{\smallskip}
\texttt{Angfluc}  & $2.8\times10^{16}$ & $1.5 \times 10^{16} $ & $4.3\times10^{12}$ & $1.5 \times 10^{14} $  & $~~9\times10^{-15}$ & $ ~~2\times 10^{-14}$ & $1\times10^{-10}$ & $2 \times 10^{-11}$ & $0.9\% \,  \tilde{\chi}_{\rm Mo}^{t0}$ &  $   5.3 \, \tilde{\chi}_{\rm Ag}^{t0}$ \\ 
\noalign{\smallskip}
\texttt{Radfluc}   & $2.8\times10^{16}$ & $1.3\times10^{16}$ & $4.0\times10^{12}$ & $1.1\times10^{14}$ & $-3\times10^{-15}$ & $-2\times10^{-15}$ & $9\times10^{-11}$ & $2\times10^{-11}$ & $0.2\% \,  \tilde{\chi}_{\rm Mo}^{t0}$ & $   7.5 \,  \tilde{\chi}_{\rm Rd}^{t0}$\\ 
\hline
\end{tabular}
\label{table:simulation}
\end{table*}

Rather than attempting to model the full diversity of magnetic-field configurations predicted by proto-neutron-star studies, we initialize the system with a magnetic field whose energy is concentrated at small spatial scales and whose spectrum peaks at a characteristic angular wavenumber $\ell_0$. At scales larger than the peak scale ($\ell<\ell_0$), we adopt the spectrum of a random vector potential. In three dimensions, this corresponds to a vector potential spectrum $E_A(\ell)\propto\ell^2$ and, consequently, a magnetic energy spectrum $E_M(\ell)\propto\ell^4$ (see Table~1 of Ref.~\cite{dehmanbrandenburg2025}). We choose $\ell_0\simeq50$, with the spectrum extending up to $\ell_{\max}=70$. For the radial wavenumber, $k_r$, we adopt values of order a few hundred, chosen to balance the fastest-growing mode of the CMI against Ohmic dissipation (see Appendix~II of Ref.~\cite{DehmanPons2025}).

The radial direction is resolved more finely than the angular directions for both physical and numerical reasons. Physically, the CME is sensitive to neutron-star microphysics, which exhibits strong radial gradients. Numerically, \MATINS is not parallelized, and the spherical harmonic decomposition required for higher angular resolution is computationally demanding. Accordingly, the crust is discretized with 200 radial grid points, allowing structures on scales of a few meters to be resolved. The angular directions are discretized using a cubed-sphere grid \citep{dehman2023} with $N_\xi = N_\eta = 47$ points per patch over six patches, corresponding to effective resolutions of $N_\theta = 94$ and $N_\varphi = 188$, and resolving angular scales of several hundred meters.

With this numerical setup, the initial magnetic field has a characteristic strength of order $10^{16}\,\mathrm{G}$ and is predominantly concentrated at small scales. The magnetic energy is close to equipartition between the poloidal and toroidal components, with a slight excess in the toroidal component. The large-scale dipolar component is limited to $\sim10^{12}\,\mathrm{G}$. The total magnetic energy of the initial configuration is $\sim10^{49}\,\mathrm{erg}$, consistent with expectations for neutron-star birth \citep{reboul2021, masada2022}. An example of the resulting initial configuration is shown in Figure~\ref{fig: bfield}, where the color scale indicates the magnetic field strength.

Since the magnetic helicity at neutron-star birth remains largely unconstrained, we explore several initial configurations spanning plausible scenarios for a newborn neutron star. This allows us to assess how the initial helicity content influences magnetic energy transfer across scales and the generation of large-scale fields. The configurations considered in this study are listed below:
\begin{enumerate}
\renewcommand{\labelenumi}{\textit{(\roman{enumi})}}
\item \texttt{Monohel} – a helical configuration with uniform helicity sign throughout the crust;
\item \texttt{Bihel} – a bi-helical configuration with opposite helicity signs in the northern and southern hemispheres;
\item \texttt{Mixhel} – a configuration with mixed helicity signs across hemispheres;
\item \texttt{Angfluc} – a configuration with small-scale angular helicity fluctuations;
\item \texttt{Radfluc} – a configuration with small-scale radial helicity fluctuations. In this case, different radial functions are used for the poloidal and toroidal components.;
\item \texttt{NoCME} – identical to the \texttt{Bihel} setup but with the CME deactivated, serving as a reference case.
\end{enumerate}
The \texttt{Bihel}, \texttt{Mixhel}, \texttt{Angfluc}, and \texttt{Radfluc} configurations are constructed to have an approximately vanishing net helicity. Details on the construction of each configuration are provided in Appendix~\ref{app: initial condition}, and the key parameters are summarized in Table~\ref{table:simulation}. The local three-dimensional distribution of the initial magnetic helicity density, normalized by its maximum absolute value, $\chi(r, \xi, \eta, t) = \AAA \cdot \BB / \left| \AAA \cdot \BB \right|_{\rm max}$, is shown in Figure~\ref{fig: initial helicity} for the \texttt{Monohel}, \texttt{Bihel}, \texttt{Mixhel}, and \texttt{Angfluc} runs. The corresponding distribution for the \texttt{Radfluc} run is shown in Figure~\ref{fig: radial fluctuations} in Appendix~\ref{app: initial condition}. This normalization is introduced solely to visualize the spatial distribution of positive and negative helicity regions and to facilitate comparisons between simulations; it does not quantify the global helicity content. Note that values of $\chi=\pm1$ correspond to the local extrema of the normalized helicity density and do not indicate a fully helical magnetic field configuration, since the normalization is performed using the maximum absolute value of the helicity density. A quantitative assessment of the helicity content of each configuration is presented in the next section.

Each setup is evolved in full crustal simulations over the first thousand years using day-long timesteps. We focus on these early stages to assess whether the CME can contribute to the formation of $10^{14}$~G dipolar magnetic fields in young magnetars, such as Swift~J1818.0--1607 ($\sim200$\,yr; \citealp{esposito2020}), on these short timescales.

\section{Results}
\label{sec: results}

\subsection{Helicity analysis}
\label{subsec: helicity analysis}

\begin{figure}
    \centering
    \includegraphics[width=\linewidth]{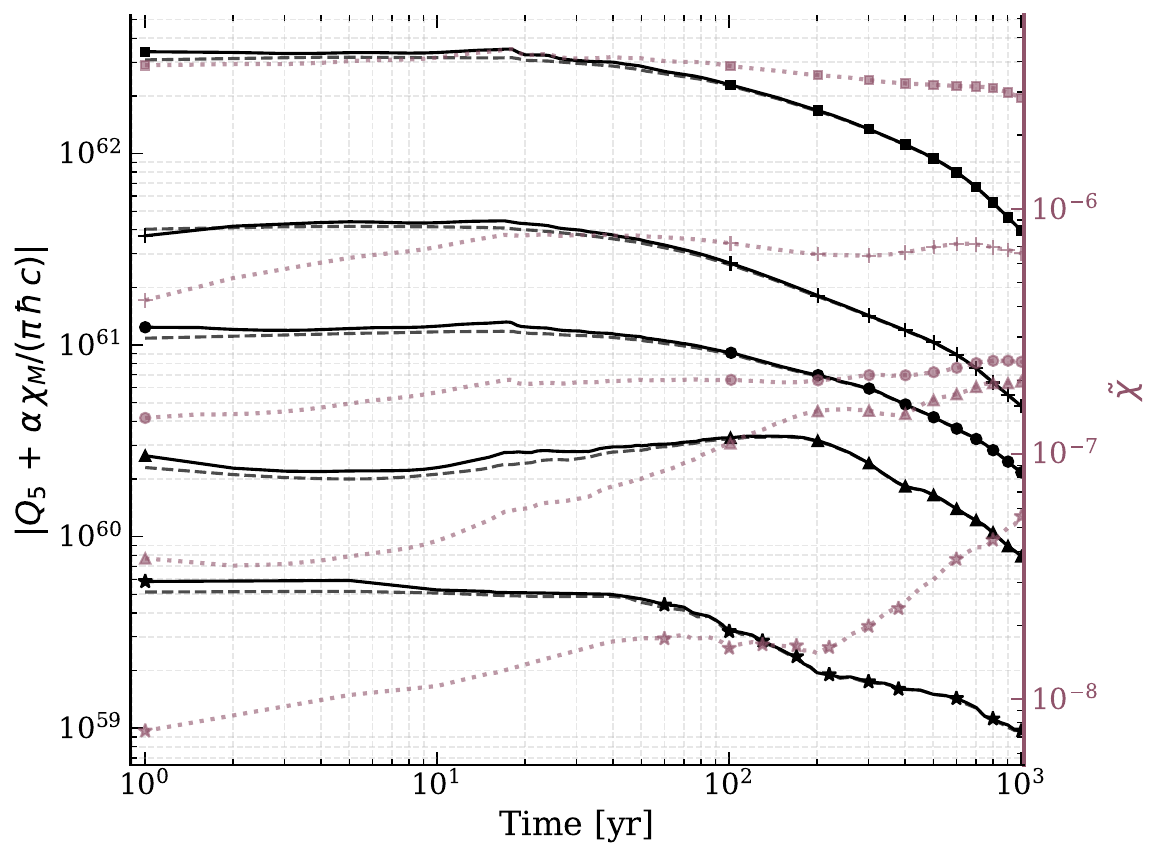}
    \caption{Evolution of helicity-related quantities. The left y-axis shows the generalized helicity (Eq.~\ref{eq: modified helicity conservation}), $|Q_5 + \frac{\alpha}{\pi\,\hbar\,c}\,\chi_M|$ (solid lines), together with the losses due to the spin-flip term, $ |\Gamma_5 \, \Delta t|$ (dashed lines). The right y-axis shows the dimensionless fractional helicity, $\tilde{\chi}$ (dots). Different simulation runs are distinguished by markers: squares (\texttt{Monohel}), circles (\texttt{Bihel}), plus signs (\texttt{Mixhel}), triangles (\texttt{Angfluc}), and stars (\texttt{Radfluc}).}
    \label{fig: conservation}
\end{figure}

The conservation of the generalized helicity (Eq.~\ref{eq: modified helicity conservation}) and its numerical verification are presented in Figure~\ref{fig: conservation} (left y-axis; black). The solid lines show the time evolution of the total helicity, $|Q_5 + \frac{\alpha}{\pi\,\hbar\,c}\,\chi_M|$, while the dashed lines represent the losses due to spin-flip processes at each time step, $|\Gamma_5 \, \Delta t|$. This quantity corresponds to the net reduction of chiral charge due to electron chirality-flipping interactions, integrated over the crust during a single time step. Different simulation runs are distinguished by markers: squares (\texttt{Monohel}), circles (\texttt{Bihel}), plus symbols (\texttt{Mixhel}), triangles (\texttt{Angfluc}), and stars (\texttt{Radfluc}).

Figure~\ref{fig: conservation} confirms that the simulations satisfy the generalized helicity conservation law (Eq.~\ref{eq: modified helicity conservation}). In particular, the evolution of the total helicity is well balanced by the spin-flip term, i.e., changes in $|Q_5 + \frac{\alpha}{\pi\,\hbar\,c}\,\chi_M|$ are closely compensated by $|\Gamma_5 \, \Delta t|$, as predicted by theory. Minor early-time discrepancies (during the first few years) arise from transient adjustments and small surface helicity fluxes ($\propto \mathbf{E} \times \mathbf{A}$) introduced by the boundary conditions. These effects decay rapidly as the system relaxes. Reducing the time step improves conservation but at a significant computational cost; nevertheless, helicity is conserved to a satisfactory level for this analysis.

The magnetic helicity is associated with an extremely small chiral charge $Q_5$, approximately 20 orders of magnitude smaller than the magnetic helicity itself (not shown here; see Figure~5 of Ref.~\citep{DehmanPons2025}). This chiral charge remains nearly constant over time. Despite its small magnitude, $Q_5$ provides the chiral asymmetry that drives the CME and, consequently, influences the magnetic-field evolution. Its value is limited by strong spin-flip damping; in the absence of spin-flip processes, $Q_5$ would grow to values approaching the total number of electrons in the crust, $\mathcal{O}(10^{55})$. Even in that limit, however, it would remain several orders of magnitude smaller than the magnetic helicity shown on the left y-axis of Figure~\ref{fig: conservation}.

Figure~\ref{fig: conservation} also shows the time evolution of the dimensionless fractional helicity, $\tilde{\chi}$ (dots, right y-axis, purple), defined as
\begin{equation}
    \tilde{\chi} = \frac{k_{\rm min} \, |\chi_M|}{2 \EM},
\end{equation}
where $k_{\rm min} \equiv 2\pi / R_{\rm crust}$ is the smallest wavenumber allowed by the domain, with $R_{\rm crust}$ representing the largest spatial scale of the system. This definition normalizes the magnetic helicity to the maximum value attainable at the system scale. By construction, $\tilde{\chi} \approx 1$ corresponds to a configuration approaching the maximum helicity allowed at the largest scale (i.e., a fully helical, nearly force-free field), whereas $\tilde{\chi} \ll 1$ indicates that the magnetic field carries little magnetic helicity at the system scale, either due to intrinsically low helicity, cancellation between regions of opposite-sign helicity in the volume integral, or helicity confined to small scales. An increase in $\tilde{\chi}$ therefore indicates a relative enhancement of magnetic helicity at the largest spatial scale. In all simulations presented here, including the \texttt{Monohel} run, $\tilde{\chi}$ remains well below unity. The initial helicity content nevertheless differs significantly between runs: \texttt{Monohel} has the largest value, followed by \texttt{Mixhel} (approximately $10\%$ of \texttt{Monohel}), \texttt{Bihel} ($\sim 3.5\%$), \texttt{Angfluc} ($\sim 1\%$), and \texttt{Radfluc} ($\sim 0.2\%$), as summarized in Table~\ref{table:simulation}.

\begin{figure*}
\includegraphics[width=1.5\linewidth,height=1.25\linewidth,keepaspectratio]{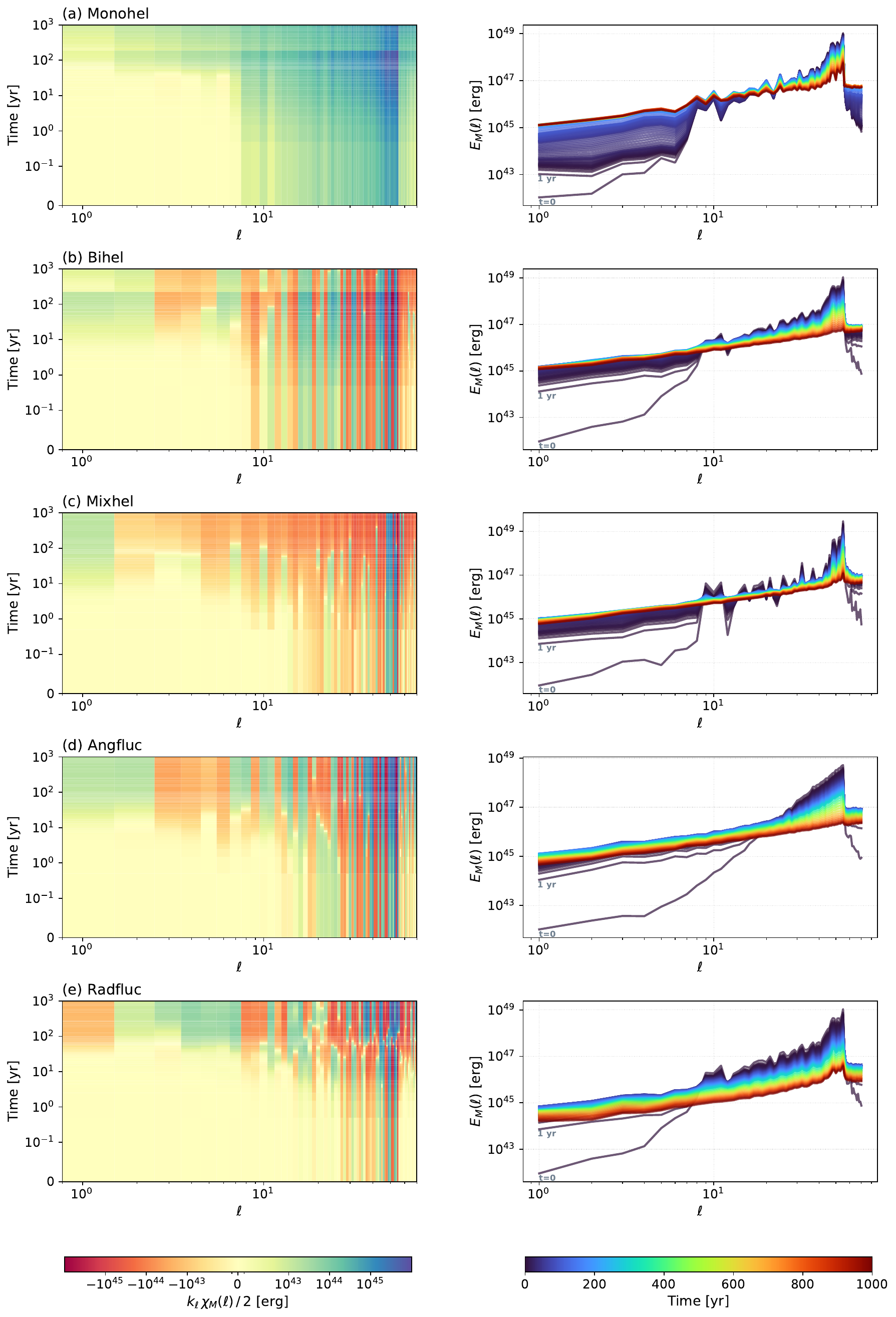}
\caption{Magnetic helicity and energy spectra for the five simulations: (a) \texttt{Monohel}, (b) \texttt{Bihel}, (c) \texttt{Mixhel}, (d) \texttt{Angfluc}, and (e) \texttt{Radfluc}. The left column shows the magnetic helicity spectrum $k_{\ell}\,\chi_M(\ell,t)/2$ as a function of spherical harmonic degree $\ell$ and time, with color indicating the helicity amplitude. The right column shows the magnetic energy spectrum $E_M(\ell)$ versus $\ell$ at different times, with color encoding time. }
\label{fig: energy helicity spectra}
\end{figure*}

The evolution of $\tilde{\chi}$ exhibits distinct behaviors across the different runs. In \texttt{Monohel}, $\tilde{\chi}$ remains approximately constant, with only a slight decrease over time. A similar trend is observed in \texttt{Mixhel} and \texttt{Bihel}, where both runs show a modest early-time increase before saturating at comparable values (see also Table~\ref{table:simulation}). This suggests that the magnetic helicity is not significantly redistributed across spatial scales and remains relatively weak at the largest scales. In contrast, the \texttt{Angfluc} and \texttt{Radfluc} runs exhibit a pronounced increase in $\tilde{\chi}$, growing by roughly an order of magnitude over time, indicating a progressive transfer of magnetic helicity toward larger spatial scales. As a result, the large-scale magnetic field becomes increasingly helical compared to both the other simulations (\texttt{Monohel}, \texttt{Bihel}, and \texttt{Mixhel}) and the smaller-scale structures within each run. This has important implications. It suggests that large-scale helical fields can themselves drive the generation of chiral asymmetry via generalized helicity conservation (Eq.~\ref{eq: modified helicity conservation}), rather than relying solely on pre-existing small-scale structures. This could, in turn, lead to faster dissipation of the large-scale fields formed through the CME, compared to standard Ohmic decay. To support this hypothesis, we further examine the magnetic helicity and energy spectra (Figure~\ref{fig: energy helicity spectra}) and the evolution of the dipolar magnetic field (Figure~\ref{fig: B mu5 evol}) in what follows.

The magnetic helicity spectrum is defined as
\begin{equation}
        \chi_M(\ell,m;t) = \frac{1}{2 \pi} \int \ell(\ell+1)  \Phi_{\ell m} \Psi_{\ell m} dr,
        \label{eq: spectral magnetic helicity}
\end{equation}
while the magnetic energy spectrum is given by
\begin{equation}
\EM(\ell,m;t) = \frac{1}{8 \pi} \int \ell(\ell+1) \bigg[  \frac{\ell(\ell+1)}{r^2} \Phi_{\ell m}^2 +\Phi^\prime_{\ell m} \,^2   +\Psi_{\ell m}^2  \bigg] dr, 
        \label{eq: energy spectrum MATINS}
\end{equation}
where $\Phi^\prime_{\ell m} = \partial \Phi_{\ell m}/\partial r$. The scalar functions $\Phi_{\ell m}$ and $\Psi_{\ell m}$ are defined in Appendix~\ref{app: initial condition}.
These expressions lead to the spectral realizability condition,
\begin{equation}
      k_{\ell}\,  \chi_M(\ell, m; t) \le  2 \, \EM(\ell,m; t),
    \label{eq: realisability l}
\end{equation}
where $k_{\ell}=\sqrt{\ell(\ell+1)}/R$ denotes the effective spherical wavenumber, with $R$ the radius of the computational domain. This condition must hold at all times for each spherical harmonic degree $\ell$. Note that in the numerical analysis (as is the case in Figure~\ref{fig: energy helicity spectra}), the spectra are obtained by summing over $m$, namely $\chi_M(\ell,t)=\sum_m \chi_M(\ell,m;t)$ and $\EM(\ell,t)=\sum_m \EM(\ell,m; t)$.

The left panels of Figure~\ref{fig: energy helicity spectra} show the evolution of $k_{\ell} \, \chi_M(\ell,t)/2$ over the first $1000$~yr of neutron star evolution as a function of spherical harmonic degree $\ell$ (horizontal axis) and time (vertical axis), with color indicating the amplitude. The right panels display the corresponding magnetic energy spectra, $\EM(\ell)$, as a function of $\ell$ at different times, with color representing the temporal evolution. Throughout the evolution, we verify that the realizability condition is satisfied, i.e., $k_{\ell}\,\chi_M(\ell,t)/2 \leq \EM(\ell,t)$ for all times and harmonic degrees $\ell$, ensuring the physical consistency of the results.

At $t=0$, $k_{\ell}\,\chi_M(\ell,t=0)/2$ is predominantly concentrated at small spatial scales, mainly in the range $10 \lesssim \ell \lesssim 55$, depending on the specific simulation. In the helicity-free runs (\texttt{Bihel}, \texttt{Mixhel}, \texttt{Angfluc}, and \texttt{Radfluc}), both positive and negative contributions are present, leading to an overall cancellation of the net magnetic helicity, whereas in the \texttt{Monohel} run the multipolar components predominantly carry a positive helicity. The large-scale structures are nearly non-helical in all simulations. Similarly, the smallest resolved scales ($60 \lesssim \ell \lesssim 70$) also carry negligible helicity. The magnetic energy (right panels) is likewise concentrated at small scales, with the initial spectrum approximately following $\EM(\ell) \propto \ell^4$, and only weak contributions from large-scale modes.

As the system evolves, the magnetic helicity remains approximately conserved, although a few individual spherical-harmonic multipoles undergo sign changes over time. Moreover, the magnetic helicity is progressively redistributed away from the initially helical small-scale structures, spreading toward both larger and smaller spatial scales that initially carried little or no helicity. This evolution is accompanied by a concurrent redistribution of magnetic energy across multipoles, following a similar pattern. The rate of this redistribution differs between the simulations. In the helicity-free runs, the transfer of both helicity and energy toward large-scale structures (e.g., dipole, $\ell=1$) proceeds rapidly, typically within about a decade (see left panels), with a noticeable redistribution of magnetic energy already occurring within the first year (see right panels). In contrast, in the helical run (\texttt{Monohel}) the same redistribution is significantly delayed: a substantial transfer of helicity to the dipolar component appears only after nearly a century, while during the first year the change in the magnetic energy-spectrum slope remains markedly smaller than in the helicity-free cases.

The slope of the magnetic energy spectrum gradually converges toward a common spectral scaling, $\EM(\ell) \propto \ell$, for all simulations, suggesting that an inverse-like cascade has taken place, with each spherical-harmonic mode saturating independently at a characteristic energy level. Despite this redistribution, small-scale structures remain dominant in the magnetic field configuration throughout the evolution. This behavior differs from the standard MHD inverse cascade, which is typically characterized by a shift of the spectral peak toward lower $\ell$ while approximately preserving the initial spectral slope \citep{brandenburg2020, dehmanbrandenburg2025}\footnote{The extent of the inverse cascade under the non-linear Hall term in neutron stars is limited by the inverse aspect ratio of the neutron star crust, $\mathcal{A} \approx 1/30$, making it ineffective in forming large-scale structures ($\ell=1$), despite its presence \citep{dehmanbrandenburg2025}.}.

An amplification of the large-scale magnetic structures ($\ell \lesssim 10$) occurs in all simulations, but it differs between runs. In the \texttt{Monohel} and \texttt{Bihel} runs, the large-scale structures, once established, remain stable over time, with energy dissipation occurring predominantly in the smaller-scale structures. By contrast, in the \texttt{Mixhel} run, the large-scale structures undergo slight dissipation at later times compared to \texttt{Monohel} and \texttt{Bihel}. A more pronounced dissipation of the formed large-scale structures is seen in the \texttt{Angfluc} run, and is even stronger in \texttt{Radfluc}. In these latter two cases, the dissipation of the large-scale structures is evidently not driven by Ohmic diffusion alone, but also by the CME. This is evident because all scales, including the smallest ones generated by the CME ($\ell \gtrsim 55$), decay at a comparable rate. In contrast, under pure Ohmic diffusion, large-scale magnetic structures would evolve more slowly, since the diffusive timescale scales as $\tau_d \equiv 1/(\eta k^2) \sim L^2/\eta$, where $k^2 = k_r^2 + k_\ell^2 \equiv -\nabla^2$, with $k_r$ and $k_\ell$ the radial and spherical wavenumbers and $L \sim 1/k$ a characteristic magnetic length scale.

We relate this behavior to the redistribution of magnetic helicity toward larger spatial scales. This redistribution is not uniform across runs, as already indicated by the evolution of $\tilde{\chi}$ in Figure~\ref{fig: conservation}. We focus on the dipolar component, both to avoid conflating contributions from multiple scales and because of its particular relevance for magnetar applications. In the \texttt{Monohel} and \texttt{Bihel} simulations, the dipole initially develops a modest level of helicity, which then decreases at later times. This is consistent with an ongoing exchange of magnetic helicity between the dipole and smaller-scale multipoles, which continuously replenishes and sustains the dipolar field.

By contrast, in the \texttt{Angfluc} and \texttt{Radfluc} runs, the dipole ultimately decays once the magnetic energy stored in the $\ell=1$ mode reaches $\sim 10^{45}$ erg, starting from an initial total magnetic energy of $\sim 10^{49}$ erg. This value represents the maximum energy attainable by the dipolar component through the CME, and it converges across all runs (see right panels of Figure~\ref{fig: energy helicity spectra}). In these two runs, the helicity contained in the dipolar component grows comparatively large at later times ($t \gtrsim 100$ yr), particularly in \texttt{Radfluc}. In this regime, the dipole acts as an intermediate reservoir: it transfers helicity to smaller scales through the CME, thereby contributing to the destabilization of the large-scale field.
These results are consistent with the evolution of $\tilde{\chi}$ (Figure~\ref{fig: conservation}). The pronounced late-time increase in $\tilde{\chi}$, particularly in \texttt{Radfluc} and \texttt{Angfluc}, indicates a progressive accumulation of helicity within the large-scale magnetic field structure — in contrast to the behavior observed in \texttt{Monohel} and \texttt{Bihel}.

\subsection{Magnetic field analysis}
\label{subsec: decay law}

\begin{figure*}[ht!]
    \centering
    \includegraphics[width=0.47\linewidth]{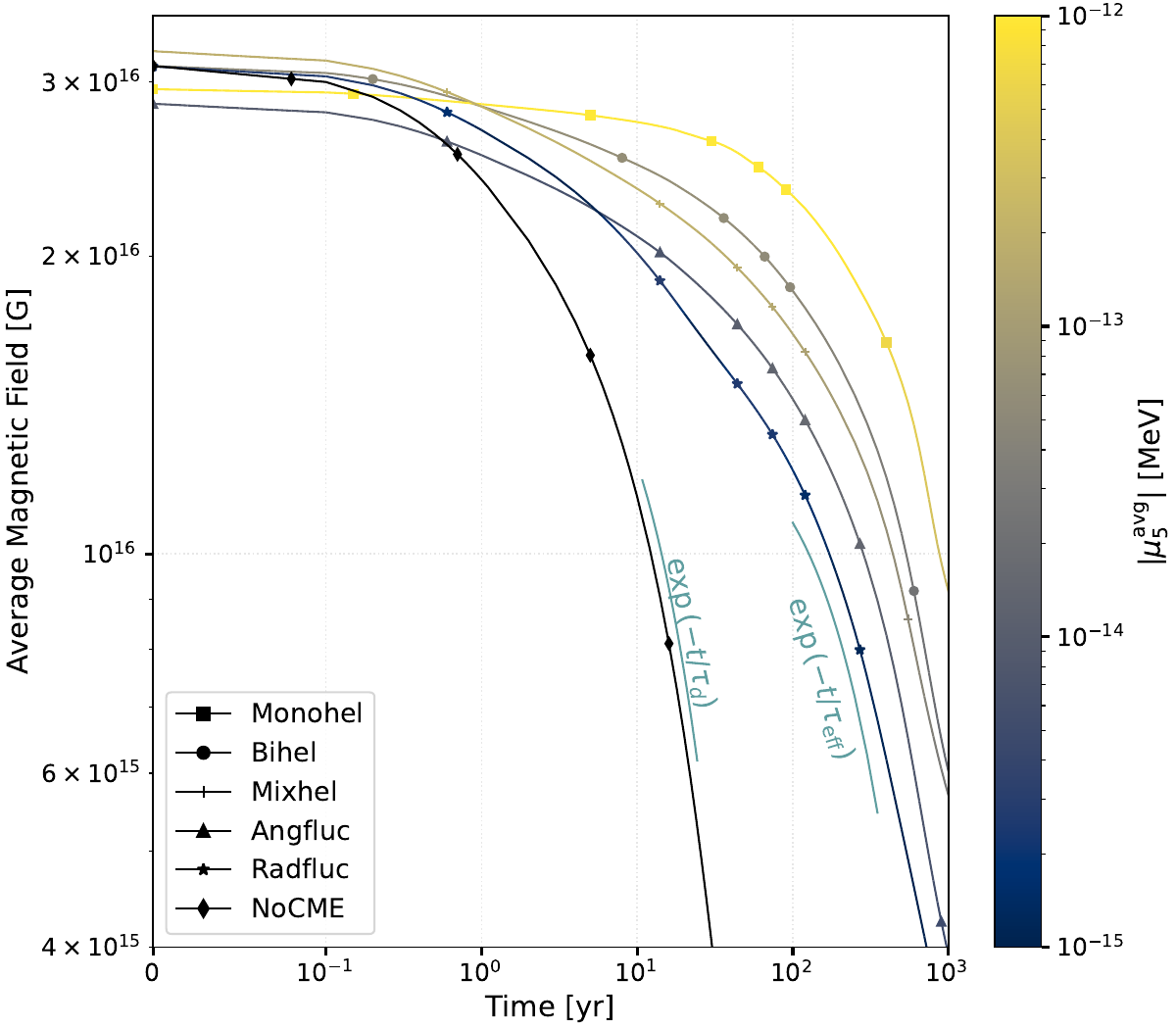}
\hspace{5pt}
\includegraphics[width=0.47\linewidth]{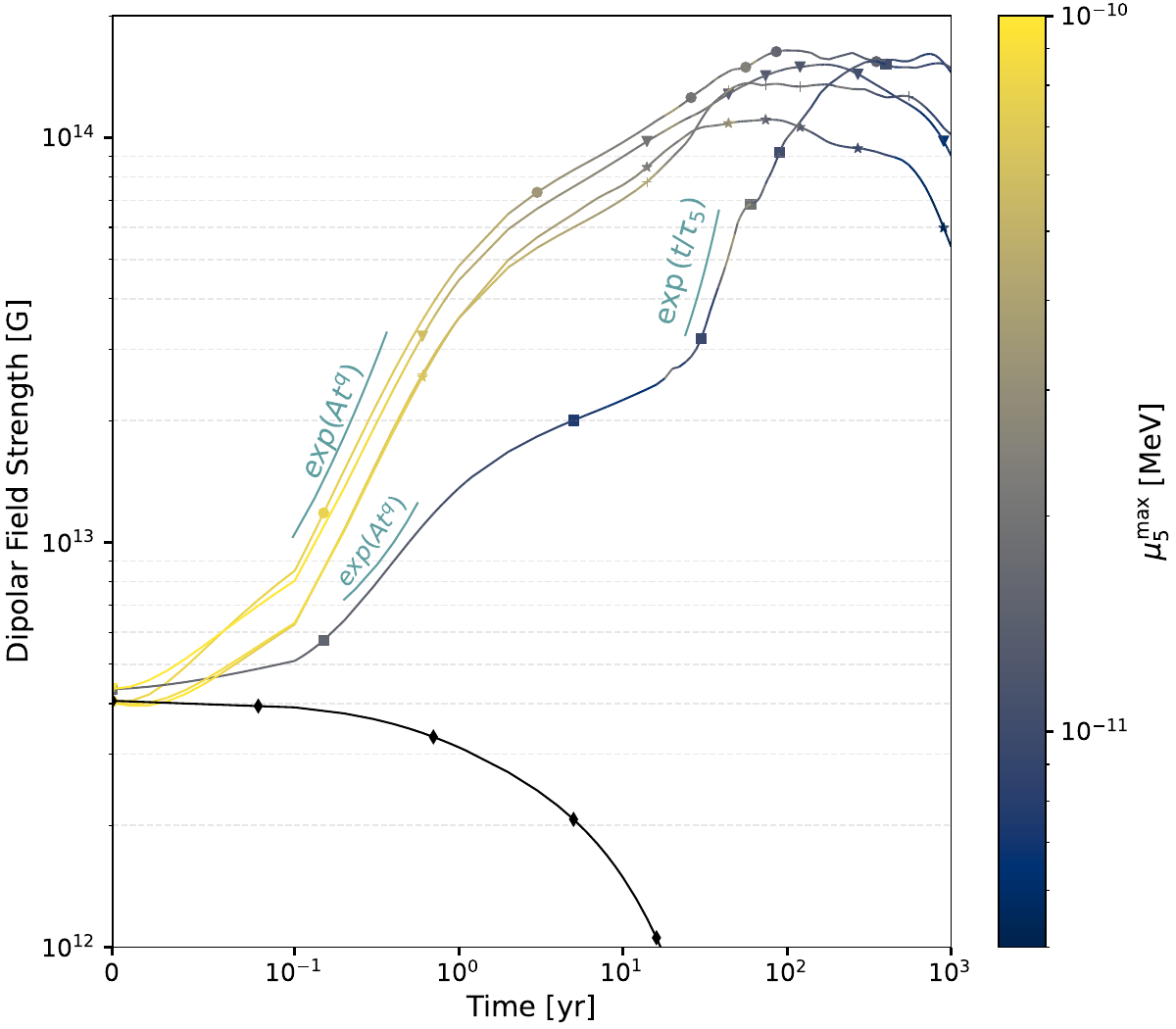}
 \caption{Magnetic field evolution over the first $10^3$ years. The left panel shows the volume-averaged magnetic field, while the right panel shows the mean dipolar magnetic field. The simulation runs are represented by solid lines with distinct markers (as in Figure~\ref{fig: conservation}). In the left panel, the colorbar corresponds to the average chiral chemical potential, $\mu_5^{\rm avg}$, while in the right panel it corresponds to the maximum chiral chemical potential, $\mu_5^{\rm max}$. The \texttt{NoCME} run is shown in black (with diamond markers), as it does not include the CME. Lines in cadetblue show fits to the decay (left panel) and growth (right panel) phases of the magnetic field.} 
\label{fig: B mu5 evol}
\end{figure*}

The time evolution of the volume-averaged magnetic field and its dipolar component is shown in Figure~\ref{fig: B mu5 evol}. The color bar in the left panel indicates $|\mu_5^{\rm avg}|$, while the right panel shows the evolution of $\mu_5^{\rm max}$. Different simulation runs are denoted by solid lines with distinct markers. The \texttt{NoCME} run is shown in black, as the CME is switched off ($\mu_5=0$).

The volume-averaged magnetic field (left panel of Figure~\ref{fig: B mu5 evol}) is initially $B_{\rm rms} \simeq 3 \times 10^{16}\,\mathrm{G}$ in all simulations and decays over time. When the CME is included --- sourced by the topology of the magnetic field --- its effect is to redistribute magnetic energy across spatial scales rather than amplify the mean field (see Figure~\ref{fig: energy helicity spectra}). As a result, the decay rate of $B_{\rm rms}$ depends on the specific simulation. In the \texttt{NoCME} run, where only Ohmic dissipation operates, the magnetic field decays exponentially as $\propto \exp(-t/\tau_{\rm d})$, with $\tau_{\rm d} \approx 20\text{--}25$~yr. All modes decay in this case, including the initially weak dipolar component. 

In the CME-active simulations, the decay departs from purely Ohmic behavior and follows $\propto \exp(-t/\tau_{\rm eff})$, with an effective dissipation timescale
\begin{equation}
\tau_{\rm eff} \approx \frac{\tau_d}{ 1 - \dfrac{|k_5|}{k}}. 
\label{eq: tau_eff}
\end{equation}
This expression reflects the competition between Ohmic dissipation and the CME. For $k \gg |k_5|$ (very small scales), $\tau_{\rm eff} \approx \tau_{\rm d}$ and the CME is negligible. For $k \gtrsim |k_5|$ (intermediate scales), dissipation is suppressed and $\tau_{\rm eff} > \tau_{\rm d}$. In the limiting case $k \to |k_5|$ (characteristic scale), $\tau_{\rm eff} \to \infty$, corresponding to complete suppression of dissipation. The regime $k < |k_5|$ (large scales), which would lead to growth, cannot be realized, since $k_5 \approx k/(1+B_{\rm sat}^2/B^2)$ implies that $k_5 < k$ at all times.

The color variation of the curves in the left panel of Figure~\ref{fig: B mu5 evol} represents different values of $|\mu_5^{\rm avg}|$. We plot the absolute value since its sign depends on the relative orientation of the electric and magnetic fields; the corresponding sign for each run is reported in Table~\ref{table:simulation}. The magnitude of $|\mu_5^{\rm avg}|$ is primarily determined by the initial magnetic-field configuration and remains approximately constant throughout the evolution, with a slight decay at later times ($t \gtrsim 200$ years). This is also consistent with the decay of the magnetic field, which follows a purely exponential form --- as expected if $\mu_5^{\rm avg}$ remains nearly constant.
The largest values are obtained in the \texttt{Monohel} run, reaching $\mu_5^{\rm avg}\sim 10^{-12}$~MeV, while the other configurations (\texttt{Mixhel}, \texttt{Bihel}, \texttt{Angfluc}, and \texttt{Radfluc}) lie in the range $\sim 10^{-13}$--$10^{-15}$~MeV. A clear trend is observed whereby larger values of $|\mu_5^{\rm avg}|$ are associated with slower magnetic-field decay, with the exception of the \texttt{Mixhel} case. This suggests that $\tau_{\rm eff}$ (see Eq.~\eqref{eq: tau_eff}), which is controlled by the chiral parameter $|k_5|$, correlates with $|\mu_5^{\rm avg}|$, with $|k_5| \equiv |k_5|^{\rm avg}$.

The time evolution of the dipolar field, shown in the right panel of Figure~\ref{fig: B mu5 evol}, differs among the various runs. In the \texttt{NoCME} case ($\mu_5=0$, black curve), the dipolar component shows no increase, consistent with the absence of the CME. It decreases from $4.0 \times 10^{12}\,\mathrm{G}$ to $1.3 \times 10^{11}\,\mathrm{G}$ over 100 years. In contrast, all other simulations show a clear growth of the dipolar field, correlated with the chiral asymmetry (see also colorbar). 

\begin{figure}[ht!]
    \centering
\includegraphics[width=\linewidth]{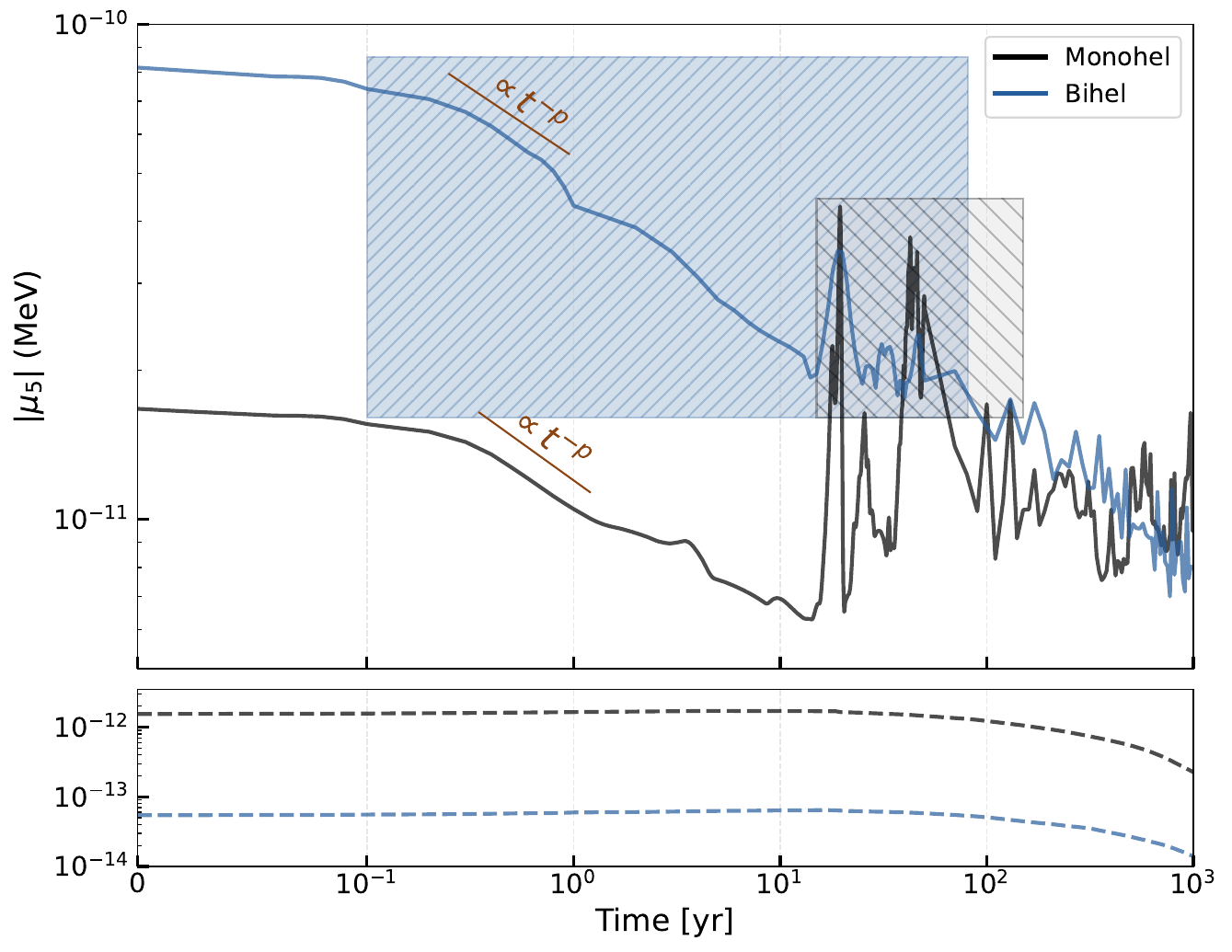}
\caption{Time evolution of the absolute value of the maximum (upper panel, solid lines) and mean (lower panel, dashed lines) of $\mu_5$ for the \texttt{Monohel} (black) and \texttt{Bihel} (blue) runs. The shaded regions indicate the time intervals during which the CMI is active in each case. Brown lines show fits to the early time decay phase of $\mu_5^{\rm max}$.}
     \label{fig: mu5 evol}
\end{figure}

The growth of the dipolar field in these CME-active runs proceeds in several stages, with clear differences between helical and helicity-free runs. During the earliest phase ($t \lesssim 0.1$~yr), all runs show only a marginal increase, slightly more pronounced in the helicity-free runs (\texttt{Bihel}, \texttt{Mixhel}, \texttt{Angfluc}, and \texttt{Radfluc}) than in \texttt{Monohel}. This is followed, in all of these runs, by a stretched-exponential growth phase, $\propto \exp(A\,t^{q})$, where $A$ and $q$ are fit parameters related to the evolution of $\mu_5$ (discussed in detail below, see Figure~\ref{fig: mu5 evol}). In each case, this phase spans two distinct time windows: a more efficient one from $\sim 0.1$ to $1$~yr, and a weaker one from $\sim 2$ to $20$~yr. This phase is markedly more pronounced in the helicity-free runs, which amplify the dipolar field to $\approx 10^{14}$~G already within the first few decades, unlike \texttt{Monohel}. At later times ($t \sim 30$ to $100$~yr), \texttt{Monohel} instead transitions to an exponential growth phase, $\propto \exp(t/\tau_5)$, with $\tau_5 \equiv (\eta\, k\, |k_5|)^{-1}$, which brings its dipolar field to a comparable strength of $\approx 10^{14}$~G. 

We interpret both the stretched-exponential phase and this subsequent exponential phase as signatures of the onset of the CMI, in both the helicity-free runs and \texttt{Monohel}. These distinct evolutionary phases highlight a clear qualitative difference between systems with and without net initial magnetic helicity. The colorbar in the right panel of Figure~\ref{fig: B mu5 evol} traces $\mu_5^{\rm max}$ over time for all CME-active runs, alongside the dipolar field growth. Across the helicity-free runs, $\mu_5^{\rm max}$ reaches values of order $10^{-10}$~MeV at early times and subsequently decreases to $\sim 10^{-11}$~MeV over several decades, broadly tracking the stretched-exponential growth phase of the dipole. In \texttt{Monohel}, $\mu_5^{\rm max}$ instead starts about an order of magnitude lower, $\approx 10^{-11}$~MeV, and evolves differently, remaining mostly constant, coinciding with its distinct growth history.

Following the onset of the CMI, the dipolar field saturates at $B \gtrsim 10^{14}$~G. In all cases considered here, the dipolar component does not exceed a few $\times 10^{14}$~G, starting from an initial magnetic field strength of $3 \times 10^{16}$~G concentrated at small scales. This indicates a robust saturation level, largely insensitive to the initial conditions explored here, a behavior also reflected in the slope of the magnetic energy spectra shown in the right column of Figure~\ref{fig: energy helicity spectra}. In particular, both the \texttt{Monohel} and \texttt{Bihel} runs reach saturation at $B \simeq 1.5 \times 10^{14}$~G, although on different timescales: after $\sim 200$~yr in the \texttt{Monohel} case and $\sim 50$~yr in the \texttt{Bihel} case. After saturation, the resulting dipolar field is stable and subsequently decays through standard Ohmic dissipation. These results indicate that the CME can generate magnetar-strength dipolar fields independently of whether the initial magnetic field is helical or helicity-free.

A distinct late-time behavior (i.e., once the $10^{14}$~G dipole has already formed) is observed in the helicity-free runs. In the \texttt{Mixhel} case, the resulting dipole is slightly weaker than in \texttt{Bihel}. It undergoes a late-time decay phase between roughly 600 and 800 years, before stabilizing at $10^{14}$~G after approximately 1000 years and subsequently evolving in a manner consistent with standard Ohmic decay. In contrast, in the \texttt{Angfluc} and \texttt{Radfluc} cases, the dipolar field enters a decay phase that is faster than expected from standard Ohmic dissipation alone; this accelerated decay sets in only once the dipolar field reaches a strength of $\approx 10^{14}$~G and develops magnetic helicity, as shown in the left panels of Figure~\ref{fig: energy helicity spectra}. In this regime, magnetic energy is transferred from the large-scale dipolar component to smaller-scale structures, leading to a weakening of the dipole. In these two cases, the resulting CME-generated dipole is therefore unstable, and this outcome depends on the initial magnetic helicity density in the system: small-scale helicity fluctuations (as in \texttt{Angfluc} and \texttt{Radfluc}) seed the formation of transient helical dipoles, which are then more susceptible to fragmentation into smaller-scale magnetic structures.

To understand the different onset of the CMI in the helicity-free versus helical case, we examine the time evolution of $\mu_5$ for the \texttt{Monohel} and \texttt{Bihel} runs, both the volume-averaged value (lower panel of Figure~\ref{fig: mu5 evol}, dashed lines) and the maximum value (upper panel, solid lines). The shaded regions in Figure~\ref{fig: mu5 evol} mark the time intervals during which the CMI actively contributes to the growth of a $\sim 10^{14}$~G dipolar field; once the dipole is established, the shading is terminated, although the CMI may still operate to sustain the field.

The volume-averaged value, $\langle \mu_5 \rangle$, decays slowly and follows a similar evolution in both runs, showing no clear correlation with the markedly different growth histories of their dipolar magnetic fields. Moreover, $\langle \mu_5 \rangle$ in \texttt{Bihel} is approximately an order of magnitude smaller than in \texttt{Monohel} --- the opposite of what would be expected if $\langle\mu_5\rangle$ controlled the CME efficiency, given that the CME is in fact more efficient in the helicity-free run (see right panel of Figure~\ref{fig: B mu5 evol}). In contrast, the maximum value, $\mu_5^{\rm max}$, differs substantially between the two runs throughout the evolution, with decay behavior that qualitatively tracks the differing growth histories of their dipolar fields. This suggests that the onset of the CMI may be governed by the magnitude of $\mu_5^{\rm max}$, although this remains speculative.

To further test this hypothesis, we start from
\begin{equation}
\frac{d \ln B}{dt} = \frac{1}{\tau_5} = \eta k |k_5^{\rm max}|,
\label{eq:tau5_rate}
\end{equation}
where 
\begin{equation}
\tau_5 \equiv \frac{1}{\eta\, k\, |k_5^{\rm max}|} \sim \frac{L}{\eta\, |k_5^{\rm max}|}.
\label{eq: tau5}
\end{equation}
We emphasize that Eq.~(\ref{eq:tau5_rate}), and consequently Eq.~(\ref{eq: tau5}), are mean-field estimates and therefore cannot accurately predict the growth rate of the dipolar field, whose evolution depends on the transfer of magnetic energy across spatial scales. A fully quantitative treatment would require a spherical-harmonic decomposition of the induction equation, including the CME term, to capture the coupling between different spherical-harmonic modes through the Clebsch--Gordan coefficients. Such an analysis is beyond the scope of the present work. Nonetheless, Eq.~(\ref{eq:tau5_rate}) predicts a specific functional form for the magnetic-field growth depending on how $|k_5^{\rm max}|$ varies in time. Assuming that $\eta$ and $k$ are time and density independent (a simplifying assumption) and that $|k_5^{\rm max}| = k_{5,0}^{\rm max}\,t^{-p}$ over a given time interval, integrating Eq.~(\ref{eq:tau5_rate}) yields exponential growth, $B(t)\propto\exp(t/\tau_5)$, in the limit $p\to0$, corresponding to constant $|k_5^{\rm max}|$. For $p=1$, the magnetic field instead grows as a power law, $B(t)\propto t^{\eta k k_{5,0}^{\rm max}}$, whereas for $p\neq1$, the solution is a stretched exponential, $B(t)\propto \exp(A\,t^{q})$, where $A=\eta k k_{5,0}^{\rm max}/(1-p)$ and $q=1-p$. 

Indeed, in both \texttt{Bihel} and \texttt{Monohel}, $\mu_5^{\rm max}$ (equivalently, $k_5^{\rm max}\equiv 4\alpha\mu_5^{\rm max}/\hbar c$) decays as a power law over the two time intervals identified above ($\sim0.1$--$1$~yr and $\sim2$--$20$~yr), with $p$ typically lying in the range $0.2$--$0.4$, depending on the run and time interval. This corresponds to $q\equiv1-p\approx0.6$--$0.8$, in agreement with the stretched-exponential growth of the dipolar magnetic field observed in both runs during these epochs (right panel of Figure~\ref{fig: B mu5 evol}). Since $\mu_5^{\rm max}$ is systematically larger in \texttt{Bihel} than in \texttt{Monohel} throughout this period, the stretched-exponential solution predicts a correspondingly stronger and faster early amplification of the magnetic field in \texttt{Bihel}, sufficient to produce a dipolar field of $\sim10^{14}$~G.

In contrast, \texttt{Monohel} reaches a comparable dipolar field strength only during a later evolutionary stage, when $\mu_5^{\rm max}$ departs from its power-law decay and instead enters a noisy plateau, fluctuating around an approximately constant value of $\mu_5^{\rm max}\approx2\times10^{-11}$~MeV (i.e., $p\rightarrow0$). This transition coincides with the onset of the exponential-growth phase identified in the right panel of Figure~\ref{fig: B mu5 evol}, indicating that this nearly constant value of $\mu_5^{\rm max}$ is sufficient to sustain exponential amplification of the dipolar magnetic field up to $\sim10^{14}$~G, comparable to the final field strength reached in the helicity-free runs, albeit through a distinct evolutionary pathway and on a longer timescale. 

Taken together, these results indicate that the onset and subsequent growth of the dipolar field are governed by $\mu_5^{\rm max}$ rather than its volume average, and a threshold value $\mu_5^{\rm max} \gtrsim 2\times 10^{-11}$~MeV appears necessary to sustain the instability. Note that this connection between the evolution of $\mu_5$ and the growth of the dipole is only approximate: in reality, $\eta$ and $k$ are time- and density-dependent, which complicates the simplified relation in Eq.~\ref{eq:tau5_rate} and makes a more precise estimate difficult to obtain analytically.

\subsection{Timescales}

\label{subsec: timescales}
\begin{figure}
    \centering
    \includegraphics[width=\linewidth]{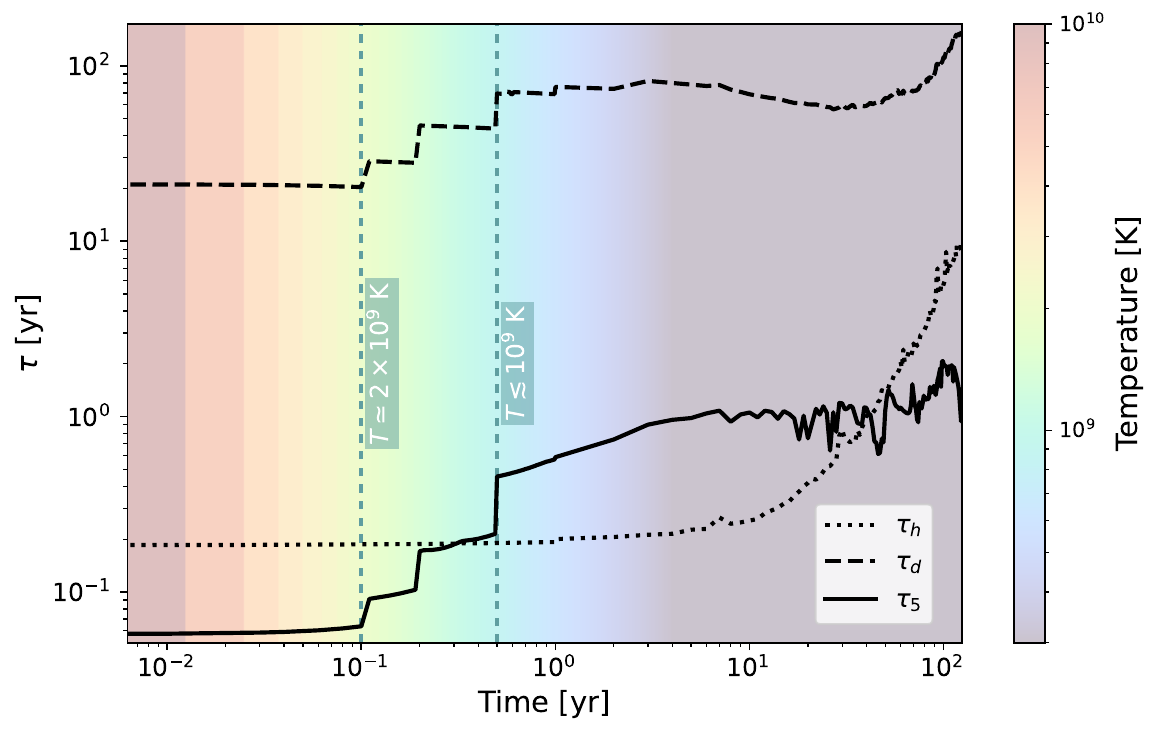}
    \caption{Time evolution of the characteristic timescales in the neutron star crust for the \texttt{Radfluc} run. The dotted, dashed, and solid black curves correspond to the Hall ($\tau_h$), Ohmic ($\tau_d$), and chiral ($\tau_5$) timescales, respectively. The color bar denotes the mean temperature evolution. Vertical green dashed lines indicate the epochs at which the temperature reaches $T \simeq 2 \times 10^9$\,K and $T \simeq 10^9$\,K.}
    \label{fig:timescales}
\end{figure}

In the present simulations, the Hall term is neglected for two main reasons. First, our goal is to isolate the effects of the CME across different magnetic field configurations and helicity contents. This becomes difficult if the Hall term is also included, since both mechanisms can transfer magnetic energy across spatial scales; combining them would therefore hinder a clear separation of their individual contributions. Second, treating both terms simultaneously poses a numerical challenge, requiring more advanced techniques than those currently implemented in the \MATINS code. To our knowledge, the combined action of the Hall term and the CME has not yet been explored in the literature.

To determine at which stage of the evolution the Hall term may become important, we examine the characteristic timescales in the neutron star crust. These include the Ohmic diffusion timescale ($\tau_d$, dashed black line), the chiral timescale ($\tau_5$, solid black line), and the Hall drift timescale ($\tau_h \approx L^2 / f_h B_{\rm rms}$, dotted black line), as shown in Figure~\ref{fig:timescales} over the first 200 years of evolution. The color bar indicates the cooling of the mean temperature, while the vertical green dashed lines mark the epochs at which the temperature reaches approximately $2 \times 10^9$\,K and $10^9$\,K, respectively. The results are shown for the \texttt{Radfluc} run.

In the early phase, the neutron star undergoes thermal relaxation, during which the temperature decreases rapidly — within less than a year — from $\sim 10^{10}$\,K to $\sim 10^9$\,K, as the crust and core approach a quasi-isothermal equilibrium \citep{potekhin2015}. This rapid cooling is reflected in the evolution of the magnetic diffusivity, which decreases as the star cools and directly impacts the Ohmic and chiral timescales, $\tau_d$ and $\tau_5$, through their dependence on $\eta$. The jumps observed in $\tau_d$ and $\tau_5$ reflect variations in the magnetic diffusivity, most pronounced at early times and gradually diminishing as the star evolves. These variations are computed self-consistently within the \MATINS code. In contrast, the Hall timescale remains largely unaffected by the temperature evolution.

The Ohmic timescale is several orders of magnitude longer (i.e., slower) than both the Hall and chiral timescales. In contrast, the Hall and chiral timescales exhibit a strong interplay. During the first few months of the neutron star's evolution, the chiral process operates on a shorter timescale than the Hall drift. As the star cools during thermal relaxation, there is a brief period ($t \approx 0.2$–$0.5~\mathrm{yr}$) in which the two timescales become comparable. Subsequently, as cooling proceeds, the ordering reverses: the Hall timescale becomes shorter than the chiral timescale by roughly a factor of two. This regime persists for a few decades, until the star reaches an age of several tens of years, at which point the Hall timescale increases significantly, while the chiral timescale remains nearly constant, with only a gradual increase due to the slow decrease of $k_5^{\rm max}$.

The divergence between the timescales at this stage results from the CME transferring magnetic energy to larger magnetic structures, which leads to a rapid increase in the characteristic magnetic length scale $L$. Since $\tau_h \propto L^2$ and $\tau_5 \propto L$, the growth of $L$ affects the Hall timescale more strongly, leading to its pronounced increase. The same scaling applies to the Ohmic timescale, $\tau_d \propto L^2/\eta$, which likewise grows progressively longer as $L$ increases.

Finally, we note that in the present evolution we do not explicitly account for changes in the characteristic magnetic length scale $L$ induced by the Hall effect. This may affect the evolution of the timescales depending on the initial conditions --- in particular, on whether the initial magnetic field supports an inverse cascade driven by the Hall term \citep{brandenburg2020,dehmanbrandenburg2025} or instead favors a direct cascade \citep{pons2007,dehman2023b}. We further emphasize that the Hall timescale does not depend directly on the magnetic helicity content or its distribution; this information is instead encoded in $k_5^{\rm max}$ and, consequently, in the chiral timescale. This distinction is important because, in regions where the CME is most effective, the magnetic field is aligned with the electric current (see Section~\ref{sec: quasi-equilibrium}). As a result, in locally force-free regions, the Hall term is not expected to significantly affect the CME-driven evolution. A fully self-consistent treatment of the coupled Hall and CME dynamics will be addressed in a follow-up study.

\section{Discussion}
\label{sec: discussion}

In the presence of an ultra-strong magnetic field, $B \gtrsim B_{\rm QED}$, as expected in magnetars, the motion of electrons perpendicular to the magnetic field becomes quantized, effectively confining their dynamics to move primarily along the magnetic field lines \citep{Kostenko2018,Thompson2020}. As the magnetic field untwists, an electric field is induced, leading to $\EE\cdot\BB > 0$. Through the chiral anomaly, electrons undergo momentum reversal along the field, effectively converting right-handed states into left-handed ones, while the resulting left-handed electrons populate higher Landau levels. Electrons in these higher-energy states are subject to chirality-flipping processes, which further reduce the chiral asymmetry (see Figure~\ref{fig:chiral anomaly}). Nevertheless, even a small chiral asymmetry, $\mu_5 \sim 10^{-11}\,\mathrm{MeV}$, can significantly modify the magnetic-field evolution by redistributing magnetic energy across spatial scales.

This mechanism has been shown to operate effectively over the long-term evolution of neutron stars hosting helical magnetic fields \citep{DehmanPons2025}. However, the magnetic helicity content at birth remains uncertain.
Standard MHD processes operating during the dynamo phase of neutron star formation preserve reflection symmetry: an initially vanishing net helicity therefore gives rise to mirror-symmetric structures of opposite handedness \citep{Woltjer1958, taylor1974, bodo2017}, allowing magnetic energy to be amplified without altering the net magnetic helicity \citep{brandenburg2005b}. By contrast, net helicity may be generated during core collapse if the CME operates during the proto-neutron star phase \citep{matsumoto2022}. This process, however, is likely suppressed by spin-flip reactions, which rapidly damp the chiral imbalance on the short dynamical timescales characteristic of proto-neutron stars \citep{grabowska2015,skoutnev2026}, thereby limiting the amount of net helicity that can be generated. Motivated by these uncertainties, we consider a range of small-scale magnetic configurations representative of newborn neutron stars, as predicted by dynamo models \citep{reboul2021}, and investigate CME-driven evolution for different initial helicity distributions, including non-helical configurations (see Figure~\ref{fig: initial helicity} and \ref{fig: radial fluctuations}).

We find that the CME efficiently generates magnetar-strength dipolar fields on decadal timescales by transferring magnetic energy from helical small-scale structures to the initially non-helical large-scale modes (see Figure~\ref{fig: energy helicity spectra}). This behavior, including the amplification of the dipole component from $\sim10^{12}$~G to magnetar-level fields of $\sim10^{14}$~G (see Figure~\ref{fig: B mu5 evol}), is largely independent of whether the initial field has a net magnetic helicity or not. In all cases considered here, the magnetic energy spectrum evolves toward a common scaling, $\EM(\ell)\propto \ell$, indicating scale-by-scale saturation of spherical harmonic modes, with the dipole component reaching energies of a few $\times10^{45}$~erg, starting from an initial total magnetic energy of order a few $\times10^{49}$~erg stored in small-scale magnetic structures.

Moreover, our results show that magnetic field configurations with vanishing net helicity are more efficient at triggering rapid dipole growth than initially helical ones. This is not a contradiction: even when the net (volume-averaged) helicity vanishes, localized helical structures are still present, and it is these structures that drive the onset of the CME by generating a residual chiral asymmetry. Indeed, the onset of the instability is primarily controlled by the maximum chiral chemical potential, $\mu_5^{\rm max}$, while the volume-averaged value plays only a subdominant role, consistent with findings in cosmological contexts \citep{Schober2021,schober2022}. In particular, the instability requires $\mu_5^{\rm max} \gtrsim \mathrm{few}\times10^{-11}~\mathrm{MeV}$ to operate efficiently in the magnetar regime, enabling the formation of strong dipolar fields of order $10^{14}$~G. This also explains why regions of opposite-sign helicity do not simply cancel in their effect on the instability: a single localized helical structure of either sign can set $\mu_5^{\rm max}$, regardless of the presence of opposite-sign helicity elsewhere in the star.

Once a $\sim 10^{14}$~G dipole is formed, its late-time evolution depends sensitively on the initial magnetic helicity density, namely the local sign and coherence of $\AAA \cdot \BB$. In some cases, these dipoles remain stable and evolve primarily through standard Ohmic decay, while in others they become unstable after acquiring sufficient helicity, leading to CME-driven dissipation and transfer of magnetic energy toward less (smaller scale) helical modes. We find that this behavior correlates with the initial magnetic helicity distribution: configurations with large coherent helicity patches (e.g., \texttt{Monohel}, \texttt{Bihel}, and \texttt{Mixhel}) produce stable, persistent dipoles, whereas highly fragmented helicity distributions (e.g., \texttt{Radfluc} and \texttt{Angfluc}) tend to generate only transient dipoles (see Figure~\ref{fig: initial helicity} and~\ref{fig: radial fluctuations} for the initial helicity density). In the latter cases, the chiral chemical potential fluctuates rapidly between positive and negative values on short spatial scales, so that no single, persistent localized helical structure survives to continuously replenish the dipole; the rapid sign reversals instead average out the local chiral asymmetry. This behavior is particularly evident in the \texttt{Radfluc} model, where helicity reversals occur on scales of $\sim 0.06$~km within a crustal thickness of $\sim 0.86$~km.

Overall, our results provide a robust pathway for magnetar formation in which initially small-scale, magnetar-strength magnetic fields can be reorganized into large-scale dipolar fields through the CME, independent of whether the magnetic field has net helicity, and without invoking a pre-existing chiral asymmetry. This reorganization requires only that the CME growth timescale be shorter than the Ohmic dissipation timescale (see Figure~\ref{fig:timescales}). The CME should therefore be systematically included in studies of magnetic field evolution in neutron-star interiors, particularly in magnetars, since its efficiency is governed by local magnetic field properties (through $k_5^{\rm max}$) rather than by the star's global magnetic helicity budget.

Finally, to address why not all neutron stars become magnetars, and to account for the observed diversity of isolated neutron-star classes, we propose the following scenario. If dynamo action operating during the proto-neutron-star stage amplifies the magnetic field to magnetar strengths on small scales, the CME acts to form strong, large-scale magnetic structures, as demonstrated in this study, regardless of the magnetic helicity content. The subsequent evolution is then primarily governed by the local magnetic helicity distribution, which determines whether the dipolar component settles into a stable configuration or becomes unstable and decays.
In the first case, the star develops a stable large-scale dipole, consistent with the observed behavior of magnetars, including their slow spin periods. In the second case, the dipolar field becomes unstable and decays, giving rise to other classes such as low-field magnetars or Central Compact Objects (CCOs), which exhibit relatively weak external (poloidal) dipolar fields despite harboring substantial internal magnetic energy \citep{dehman2023b,igoshev2025}.

Moreover, the CME scenario discussed here may also naturally account for the puzzling properties of CCOs, several of which are difficult to explain without invoking a strong, coherent internal magnetic field, most plausibly in the form of a buried toroidal component. In particular, CCOs exhibit nearly pure thermal X-ray emission with an unusually high pulsed fraction, reaching $\sim60$--$80\%$ in some sources (e.g., CXOU J185238.6+004020 \citep{Bogdanov2014}). This cannot be readily explained by a nearly uniform surface temperature, a weak dipolar field, or standard neutron-star atmosphere models; instead, it points to highly localized hot regions and strongly anisotropic heat transport through the crust. Phase-resolved spectroscopy of CCOs, including 1E 1207.4$-$5209 \citep{deluca2004}, RX J0822$-$4300 \citep{Gotthelf_2009}, and CXOU J185238.6+004020 \citep{Bogdanov2014}, further requires multiple thermal components and small emitting areas, implying strongly anisotropic surface temperature distributions that are difficult to reproduce without a strong internal field channeling the heat flux. CCOs may therefore require a CME-activation process similar to the one described in this study to explain their properties, with their external poloidal dipole further weakened or buried as a result of accretion \citep{vigano2012}.

By contrast, if the magnetic field obtained at birth is intrinsically weak, as expected for typical rotation-powered pulsars, the CME is not triggered, preventing any substantial redistribution of magnetic energy. These results highlight the importance of the magnetic helicity distribution established during the proto-neutron-star dynamo, a property that has received considerably less attention than the magnetic-field configuration itself. Both the magnetic-field structure and its helicity density should be regarded as fundamental initial conditions that determine the subsequent magnetic evolution and, ultimately, the observational class of the neutron star. Future studies should incorporate additional physical processes, in particular the coupled evolution of the CME and the Hall effect, to build a more complete theory of neutron-star magnetic-field evolution.

\section*{Data availability}
The \MATINS code \cite{dehman2023,dehman2023b,ascenzi2024} is publicly available at \href{https://github.com/ice-csic-astroexotic/MATINS}{https://github.com/ice-csic-astroexotic/MATINS}. An extended version of the code, including the effects of the chiral anomalies considered here and described in \citep{DehmanPons2025}, can be provided upon reasonable request. The reduced simulation input and output data are available on Zenodo at \href{https://doi.org/10.5281/zenodo.21112999}{https://doi.org/10.5281/zenodo.21112999}.

\begin{acknowledgements}
CD thanks the anonymous referee for a careful and constructive report that improved the manuscript. CD also thanks M. Reinhardt, N. Yamamoto, F. Coti Zelati, A. Brandenburg, J. Schober, J. Pons, D. Viganò, and S. Ascenzi for brief but valuable exchanges. CD is supported by the Ministerio de Ciencia, Innovación y Universidades (JDC2023-052227-I), co-funded by AEI (MCIN/AEI/10.13039/501100011033), the FSE+, and the Universidad de Alicante. CD acknowledges support from the Conselleria d'Educació, Cultura, Universitats i Ocupació de la Generalitat Valenciana (grant CIPROM/2022/13), and the allocation of computing resources provided by the Swedish National Allocations Committee at the Center for Parallel Computers at the Royal Institute of Technology in Stockholm (Sweden). CD made use of the publicly available version of the 3D magneto-thermal code \MATINS, funded by the European Research Council via the ERC Consolidator grant ``MAGNESIA'' (No.\ 817661; PI: N.\ Rea).
\end{acknowledgements}

\appendix

\section{Initial magnetic field and helicity distribution}
\label{app: initial condition}

To define the initial magnetic field configurations described in Section~\ref{sec: numerical setup}, we decompose the magnetic field into poloidal ($\BB_p$) and toroidal ($\BB_t$) components \citep{chandrasekhar1957}:
\begin{equation}
\BB = \BB_p + \BB_t,
\end{equation}
\begin{eqnarray}
  &&  \BB_t = - \boldsymbol{r} \times \Bnabla \Psi, \quad \quad   \AAA_t = - \rr \times \Bnabla \Phi, \nonumber \\
  &&   \BB_p = \Bnabla \times \AAA_t = - \boldsymbol{r} \Delta \Phi + \Bnabla \frac{\partial }{\partial r} \left(r \Phi \right),
\label{eq: pol tor relations}
\end{eqnarray}
where $\AAA_t$ is the toroidal vector potential. The two scalar functions $\Phi(\rr,t)$ and $\Psi(\rr,t)$ uniquely define the poloidal and toroidal components, respectively. Next, we expand the scalar functions in spherical harmonics \cite{KR80}: 
\begin{eqnarray}
    \Phi(t,r,\theta,\phi) &=& \frac{1}{r}\sum_{\ell m} \Phi_{\ell m}(r,t) Y_{\ell m}(\theta,\phi),  \nonumber\\
    \Psi(t,r,\theta,\phi)  &=& \frac{1}{r}\sum_{\ell m} \Psi_{\ell m}(r,t) Y_{\ell m}(\theta,\phi), 
       \label{eq: phi and Psi scalar functions}
  \end{eqnarray}
where $\ell = 1, 2, \ldots$ denotes the multipole degree and $m = -\ell, \ldots, \ell$ the azimuthal order. 

We define the radial dependence of the poloidal and toroidal scalar functions at $t=0$ as \citep{aguilera2008}
\begin{eqnarray}
\Phi_{\ell m}(r) &=& \Phi^0_{\ell m} \, k^p_r \, r \, \bigl[a(k^p_r r) 
    + \tan(k_r^p R) \, b(k^p_r r)\bigr], \quad \nonumber \\
\Psi_{\ell m}(r) &=& \Psi^0_{\ell m} \, k^t_r \, r \, \bigl[a(k^t_r r) 
    + \tan(k_r^t R) \, b(k^t_r r)\bigr], \quad
\label{eq: funa function} 
\end{eqnarray}
where $a(x) \equiv j_1(x)$ and $b(x) \equiv n_1(x)$ are the spherical Bessel functions of the first and second kind of order 1, respectively,
\begin{eqnarray}
a(x) = \frac{\sin x}{x^2} - \frac{\cos x}{x},  \quad 
b(x) = -\frac{\cos x}{x^2} - \frac{\sin x}{x},
\end{eqnarray}
determined by imposing the inner and outer boundary conditions. The real radial wavenumbers $k_r^p$ and $k_r^t$ are taken in the range 
$k_r \approx 400$--$450\,\mathrm{km}^{-1}$, chosen to balance the fastest-growing CME modes against Ohmic dissipation (see also Figure~3 of Ref.~\cite{DehmanPons2025}). This choice is particularly important in the radial direction, where both processes are highly sensitive to microphysical properties that vary strongly with density. Note that this choice for the toroidal radial scalar function initially introduces a tiny current sheet at the stellar surface, and thus does not perfectly satisfy the potential boundary conditions imposed at $t=0$; however, the current sheet vanishes after a few timesteps of enforcing these conditions.

The normalization coefficients $\Phi^0_{\ell m}$ and $\Psi^0_{\ell m}$ specify the amplitude and phase of each $(\ell,m)$ mode, controlling the distribution of magnetic energy across spatial scales. At scales larger than the peak scale ($\ell<\ell_0$; here we choose $\ell_0\simeq50$, with $\ell_{\max}=70$), we adopt the spectrum of a random vector potential. 
In three dimensions, this corresponds to a vector potential spectrum $E_A(\ell)\propto\ell^2$ and, consequently, a magnetic energy spectrum $E_M(\ell)\propto\ell^4$ in the subinertial range (see also Table~1 of Ref.~\cite{dehmanbrandenburg2025}). This behavior, commonly referred to as a causal spectrum in cosmology \citep{DC03}. It means that no point is correlated with any other, but the field is additionally divergence-free. 

The local magnetic helicity density is governed by the sign and relative magnitude of these coefficients for each mode $(\ell, m)$: the same sign of $\Phi^0_{\ell m}$ and $\Psi^0_{\ell m}$ implies positive local helicity, while opposite signs imply negative local helicity (see also Eq.~\ref{eq: spectral magnetic helicity}). Consequently, the identical functional forms of $\Phi_{\ell m}$ and $\Psi_{\ell m}$ in Eq.~\eqref{eq: funa function} do not predetermine the helicity of the field; varying $\Phi^0_{\ell m}$ and $\Psi^0_{\ell m}$ independently is sufficient to construct any desired helicity distribution. Additionally, introducing distinct wavenumbers $k_r^p \neq k_r^t$ — as in the \texttt{Radfluc} run — imprints small-scale fluctuations in the helicity along the radial direction. The specific choices adopted for each run are described below; the corresponding data are publicly available on Zenodo \citep{zenodo_data}.

For the \texttt{Monohel} configuration, in which an initial net magnetic helicity is imposed, the poloidal and toroidal components are intrinsically linked by construction \citep{dehmanbrandenburg2025}. Accordingly, $\Phi^0_{\ell m}$ are taken from Ref.~\cite{dehmanbrandenburg2025} (see Section III), where the weight of the multipoles was built from a random vector potential as described above. We set $k_r^p = k_r^t$ and define $\Psi^0_{\ell m}$ as
\begin{equation}
\Psi^0_{\ell m} = k_{\ell} \, \Phi^0_{\ell m},
\label{eq: Psi = k Phi}
\end{equation}
where $k_{\ell}$ is the angular wavenumber. For a maximally helical configuration, one instead sets $\Psi^0_{\ell m} = (k_r + k_{\ell})\,\Phi^0_{\ell m}$, which includes both the radial and angular wavenumber contributions. Note that $\Phi^0_{\ell m}$ and $\Psi^0_{\ell m}$ have the same sign for each mode $(\ell,m)$, so that the local magnetic 
helicity density $\mathbf{A} \cdot \mathbf{B} \propto k\,\Phi\,\Psi$ is positive throughout the entire volume, resulting in a net positive magnetic helicity $\chi_M$.

\begin{figure}
    \centering
    \includegraphics[width=0.7\linewidth]{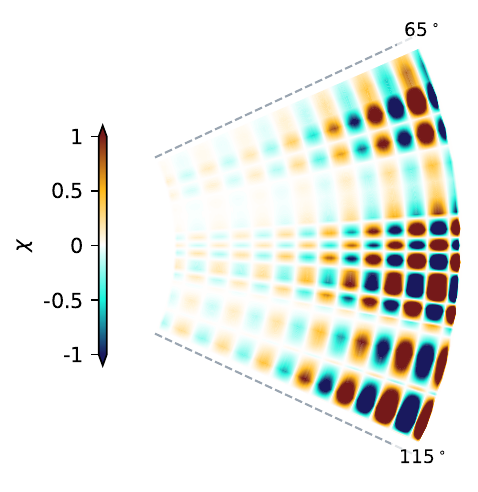}
\caption{Meridional profile of the initial local magnetic helicity $\chi$, normalized by its maximum absolute value, for the \texttt{Radfluc} run in the angular interval $65^\circ \leq \theta \leq 115^\circ$. The longitude is fixed at $0^\circ$. The crustal region is shown with a radial magnification factor of 8 for clarity.}
    \label{fig: radial fluctuations}
\end{figure}

To avoid imposing an initial net magnetic helicity by construction, we define $\Phi^0_{\ell m}$ and $\Psi^0_{\ell m}$ independently — that is, without introducing correlations between them — for the \texttt{Bihel}, \texttt{Mixhel}, and \texttt{Angfluc} runs. We also ensure that the characteristic angular scale of magnetic helicity structures decreases across these runs by progressively favoring higher spherical-harmonic modes (see the left panels of Figure~\ref{fig: energy helicity spectra}). In these three simulations, we set $k_r^p = k_r^t$, so that the same helicity distribution is reproduced at each radial layer. In the \texttt{Bihel} run, we use the same $\Psi^0_{\ell m}$ as in the \texttt{Monohel} run, but this time we select a different set of weights for the multipoles of $\Phi^0_{\ell m}$, also constructed to ensure a magnetic energy spectrum $E_M(\ell)\propto\ell^4$ in the subinertial range, as described above. At each mode $(\ell,m)$, $\Phi^0_{\ell m}$ has a different sign and magnitude compared to $\Psi^0_{\ell m}$, and the combination of $\Phi^0_{\ell m}$ and $\Psi^0_{\ell m}$ is chosen such that $\AAA \cdot \BB \propto k\,\Phi\,\Psi$ is negative in the northern hemisphere and positive in the southern hemisphere, producing a mirror-symmetric configuration with negligible net helicity. Similar logic is followed for \texttt{Mixhel}, where again the radial toroidal scalar function $\Psi^0_{\ell m}$ is the same as in the \texttt{Monohel} run, and we select a new set of multipoles for the radial poloidal scalar function $\Phi^0_{\ell m}$, independent of $\Psi^0_{\ell m}$ for each mode $(\ell,m)$, to ensure that $\AAA \cdot \BB$ is both positive and negative across each hemisphere, resulting in an almost vanishing net helicity.

In the \texttt{Angfluc} simulation, the amplitude and sign of $\Phi^0_{\ell m}$ and $\Psi^0_{\ell m}$ are assigned randomly and independently for each mode $(\ell,m)$: each coefficient is drawn from a uniform distribution in $[-1,1]$ and scaled by an $\ell$-dependent factor, chosen so that the resulting magnetic energy spectrum follows $E_M(\ell)\propto\ell^4$ in the subinertial range, consistent with the scaling adopted for the other configurations. This procedure results in $\AAA \cdot \BB$ being both positive and negative across each hemisphere, with pronounced local helicity fluctuations at each radial layer and a vanishing net helicity (see Figures~\ref{fig: initial helicity} and \ref{fig: energy helicity spectra}).

For the \texttt{Radfluc} run, we adopt the same $\Phi^0_{\ell m}$ and $\Psi^0_{\ell m}$ as in the \texttt{Bihel} run, but assign different radial wavenumbers to the poloidal and toroidal scalar functions, with $k_r^p \approx 400\,\mathrm{km}^{-1}$ and $k_r^t \approx 450\,\mathrm{km}^{-1}$. This introduces small-scale, localized helicity fluctuations in the radial direction while maintaining an approximately vanishing net helicity. Figure~\ref{fig: radial fluctuations} shows the meridional profile of the normalized magnetic helicity density, $\chi(r, \theta, \phi, t_0)$, in the angular interval $65^\circ \leq \theta \leq 115^\circ$, highlighting strong spatial variations of helicity density across the crust.

Finally, we employ curl operators within a finite-volume scheme adapted to cubed-sphere coordinates~\citep{dehman2023} and compute the magnetic field components from the poloidal and toroidal scalar functions using Eq.~\eqref{eq: pol tor relations}. This procedure guarantees an initial magnetic field that is divergence-free (to machine precision) and free of coordinate singularities.

\bibliography{apssamp}
\bibliographystyle{apsrev4-1}
\newpage
\onecolumngrid
\makeatletter

\end{document}